\documentclass[aps,pre,twocolumn,superscriptaddress]{revtex4}

\usepackage{color}
\usepackage{amsmath}    
\usepackage{amssymb} 
\usepackage{amsfonts}   
\usepackage{graphicx}

\usepackage{bbm}
\graphicspath{{./appendixgr/}}

\newcommand{\beq}{\begin{eqnarray}}
\newcommand{\eeq}{\end{eqnarray}}
\newcommand{\<}{\langle}
\renewcommand{\>}{\rangle} 
\newcommand{\abs}[1]{|#1|}

\begin{document}
\title{Fluctuating fitness shapes the clone size distribution of immune repertoires}

\author{Jonathan Desponds}
\affiliation{Laboratoire de physique th\'eorique,
    CNRS, UPMC and \'Ecole normale sup\'erieure, 24, rue Lhomond,
    75005 Paris, France}
\author{Thierry Mora}
\affiliation{Laboratoire de physique statistique, CNRS, UPMC and \'Ecole normale sup\'erieure, 24, rue Lhomond, 75005 Paris, France}
\author{Aleksandra M. Walczak}
\affiliation{Laboratoire de physique th\'eorique,
    CNRS, UPMC and \'Ecole normale sup\'erieure, 24, rue Lhomond,
    75005 Paris, France}

\begin{abstract}

The adaptive immune system relies on the diversity of receptors expressed on the surface of B and T-cells to protect the organism from a vast amount of pathogenic threats. The
proliferation and degradation dynamics of different cell types  (B cells, T cells, naive, memory) is governed by a variety of antigenic and environmental signals, yet the observed clone
sizes follow a universal power law distribution. Guided by this reproducibility we propose effective models of somatic evolution where cell fate depends on an effective fitness. This fitness is determined by growth factors acting either on clones of cells with the same receptor responding to specific antigens, or directly on single cells with no regards for clones. We identify fluctuations in the fitness acting specifically on clones as the essential ingredient leading to the observed distributions. Combining our models with experiments we characterize the scale of fluctuations in antigenic environments and we provide tools to identify the relevant growth signals in different tissues and organisms. Our results generalize to any evolving population in a fluctuating environment.

\end{abstract}

\maketitle

\section{Introduction}

Antigen-specific receptors expressed on the membrane of B and T cells (BCRs and TCRs
) recognize pathogens and initiate the adaptive immune response \cite{Janeway}. 
An efficient response 
relies on the large diversity of receptors that 
 protect us against the many pathogens in the environment. The adaptive repertoire diversity is maintained from a source of newly generated cells, each expressing a unique receptor.  These progenitor cells later divide or die, and their offspring make up clones of cells 
that share a common receptor. The sizes of clones vary, as they depend on the particular history of cell divisions and deaths in the clone. The clone size distribution thus bears signatures of the challenges faced by the  adaptive system. Understanding the form of the clone size distribution in healthy individuals is an important step in the characterization of the antigenic recognition process 
and the functioning of the adaptive immune system. It also presents an important starting point for describing statistical deviations seen in individuals with compromised immune responses.

High throughput sequencing experiments in different cell types and species have allowed for the quantification of clone sizes and their distributions \cite{quake-2009,friedman-2012, chain-2014, greenberg-2012, robins-2011, carlson-2009, mamedov-2014, Warren2011}. Generically,
clone size distributions exhibit heavy tails  \cite{quake-2009, mora2010, Zarnitsyna2013, Warren2011}---the number of clones of a given size is 
well described by a power law over a wide range of clone sizes (see Fig.~\ref{firstgraph}A-B). Such universal behavior for a variety of cells types in different species (B cells, T cells, naive cells, effector cells in fish, mice, and humans) suggests common underlying rules for growth, death and homeostatic control of adaptive repertoires. Previous work has described the specifics of immune dynamics for a certain cell type \cite{perelson-2001, Stirk2010}, or a certain signaling pathway, using detailed mechanistic models \cite{Almeida2012, Hapuarachchi2013, Reynolds2012}.
It remains unclear, however, what essential features of these dynamics may lead to the observed power-law distributions, and what are the key biological parameters of the repertoire dynamics that govern its behavior. 

The wide range and types of interactions that influence a B or T cell fate happen in a complex, dynamical environment with inhomogeneous spatial distributions. They are difficult to measure  {\em in vivo}, making  their quantitative characterization elusive. 
To overcome this, we describe the effective interaction between the immune cells and their environment as a stochastic process governed by only a few relevant parameters. This effective description is consistent with the idea, common in physics, that the apparent universality of the observed distribution is underlied by broad model universality classes. Since the heavy tail behavior exists in both B and T cells, we consider general properties that are common to both cell types, and ignore hypermutations for simplicity.

All cells proliferate and die depending on the strength of antigenic and cytokine signals they receive from the environment, which together determine their net growth rate. This effective fitness that fluctuates in time is central to our 
description. We find that its general properties determine the form of the clone size distribution.  
Two broad classes of models are distinguished, according to whether these fitness fluctuations are clone specific (mediated by their specific BCR or TCR) or cell specific (mediated by phenotypic fluctuations such as the number of cytokine receptors). We identify the models that are compatible with the experimentally observed distributions of clone sizes. These distributions do not depend on the detailed mechanisms of cell signaling and growth, but rather emerge as a result of self-organisation, with no need for fine-tuned interactions.

\section{Results}

\subsection{Clone dynamics in a fluctuating antigenic landscape}

The fate of the cells of the adaptive immune system depends on a variety of clone-specific stimulations. The recognition of pathogens triggers large events of fast clone proliferation followed by a relative decay, with some cells being stored as memory cells to fend off future infections. 
Naive cells, which have not yet recognized an antigen, do not usually undergo such extreme events of proliferation and death, but their survival relies on short binding events (called ``tickling'') to antigens that are natural to the organism (self-proteins) \cite{Troy2003, mak2006immune}. 
Because receptors are conserved throughout the whole clone (ignoring B cell hypermutations), clones that are better at recognizing self-antigens and pathogens will on average grow to larger populations than bad binders. By analogy to Darwinian evolution, they are ``fitter'' in their local, time-varying environment \cite{DeBoer1994, perelson-2001, Freitas1995}.

We denote by $a_j(t)$ the overall concentration of an antigen $j$ as a function of time. We assume that after its introduction at a random time $t_j$, this concentration decays exponentially with a characteristic lifetime of antigens $\lambda^{-1}$, $a_j(t)=a_{j,0}e^{-\lambda (t-t_j)}$ as pathogens are cleared out of the organism, either passively or through the action of the immune response. Lymphocyte receptors are specific to certain antigens, but this specificity is degenerate, a phenomenon refered to as cross-reactivity or poly-specificity. The extend to which a lymphocyte expressing receptor $i$ interacts with antigen $j$ is encoded in the cross-reactivity function $K_{ij}$, which is zero if $i$ and $j$ do not interact, or a positive number drawn from a distribution to be specified, if they do. In general, interactions between lymphocytes and antigens effectively promote growth and suppress cell death, but for simplicity we can assume that the effect is restricted to the division rate.
In a linear approximation, this influence is proportional to $\sum_j K_{ij}a_j(t)$, {\em i.e.} the combined effect of all antigens $j$ for which clone $i$ is specific. This leads to the following dynamics for the evolution of the size $C_i$ of clone $i$:
\beq\label{eq:simul}
\frac{dC_i}{dt}=\left(\nu+\sum_j K_{ij}a_j(t)-\mu\right)C_i+B\xi_i(t), 
\eeq
where $\nu$ and $\mu$ are the basal division and death rates, and 
where $B\xi_i(t)$ is a birth-death noise of intensity $B^2=(\nu+\sum_j K_{ij}a_j(t)+\mu)C_i$, with $\xi_i(t)$ a unit Gaussian white noise (see Appendix \ref{appbirthdeath} for details about birth death noise).

New clones, with a small typical initial size $C_0$, are constantly produced and released into the periphery with rate $s_C$. For example, a number of the order of $s_C=10^8$ new T-cells are output by the thymus daily in humans \cite{Bains2009}. Since the total number of T cells is of the order of $10^{11}$, this means that cells die with an average rate of $10^{-3}$ days${}^{-1}$ in homeostatic conditions \cite{Bains2009}.
Because the probability of rearranging the exact same receptor independently is very low ($< 10^{-10}$) \cite{murugan-2012}, we assume that each new clone is unique and comes with its own set of cross-reactivity coefficients $K_{ij}$.  
Assuming a rate $s_A$ of new antigens,
the average net growth rate in Eq.~\ref{eq:simul} is $f_0=\nu+\<a_{j,0}\> \langle K\rangle s_{\rm A} \lambda^{-1}-\mu <0$, and the stationary number of clones should fluctuate around $N_C\approx s_{C}|f_0|^{-1}$ clones.
This is just an average however, and treating each clone independently may lead to large variations in the total number of cells ({\em i.e.} the sum of sizes of all clones). 
To maintain a constant population size, clones compete with each other for specific resources (pathogens or self-antigens) and homeostatic control can be maintained by a global resource such as Interleukin 7 or Interleukin 2. Here we do not model this homeostatic control explicitly, but instead assume that the division and death rates $\nu,\mu$ are tuned to achieve a given repertoire size. We verified that adding an explicit homeostatic control did not affect our results (see Fig.~\ref{homeocon} and Appendix \ref{homeostasis}). 

\begin{figure}
\begin{center}
\noindent\includegraphics[width=\linewidth]{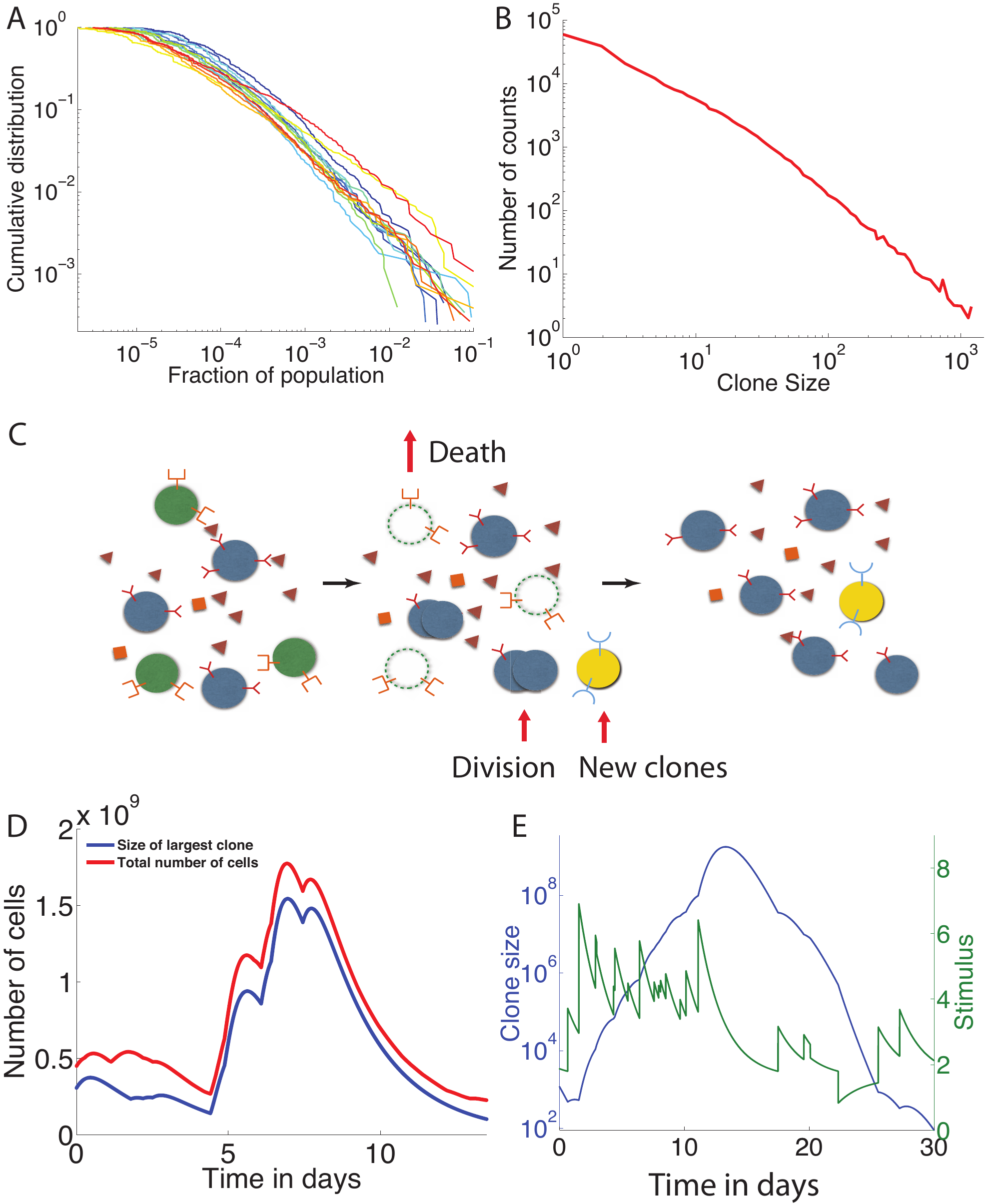}
\caption{Experimental clone size distributions have heavy tails. {\bf A.} B cell zebrafish experimental cumulative clone size distribution  for fourteen fish as a function of the fraction of the population  occupied by that clone from data in Weinstein et al. \cite{quake-2009}. 
{\bf B.} Clone size distribution for murine T-cells from Zarnitsyna et al. \cite{Zarnitsyna2013} 
(data plotted as presented in original paper). {\bf C.} 
The dynamics of adaptive immune cells include specific interactions with antigens that promote division and prevent cell death. New cells are introduced from the thymus or bone marrow with novel, unique receptors. Division, death and thymic or bone marrow output on average balance each other to create a steady state population. {\bf D-E.} Example trajectories from simulations of the immune cell population dynamics in Eq.~\ref{eq:simul}. The total number of cells (D) shows large variations
after an exceptional event of a large pathogenic invasion. One or a few cells that react to that specific antigen grow up to a macroscopic portion of the total population, and then decrease back to normal sizes after the invasion.
A typical clone size trajectory along with its pathogenic stimulation $\sum_j K_{ij}a_j(t)$ shows the coupling between clone growth and variations of the antigenic environment (E). Parameters used: $s_{\rm C}=2000$ day${}^{-1}$ , $C_0=2$, $s_{\rm A}=1.96\cdot 10^{7}$ day${}^{-1}$, $a_{j,0}=a_0=1$, $\lambda=2$  day${}^{-1}$, $p=10^{-7}$, $\nu=0.98$ day${}^{-1}$, $\mu=1.18$ day${}^{-1}$. \label{firstgraph} }
\end{center}
\end{figure}

We simulated the dynamics of a population of clones interacting with a large population of antigens. Each antigen interacts with each present clone with probability $p=10^{-7}$, and with strength $K_{ij}$ drawn from a Gaussian distribution of mean 1 and variance 1 (truncated to positive values). A typical trajectory of the antigenic stimulation undergone by a given clone, $\sum_j K_{ij} a_j$, is shown in Fig.~\ref{firstgraph}E (green curve), and shows how clone growth tracks the variations of the antigenic environment. When the stimulation is particularly strong, the model recapitulates the typical behaviour experimentally observed at the population level following a pathogenic invasion \cite{Murali-Krishna1998, Kaech2002}, as illustrated in Fig.~\ref{firstgraph}D: the population of a clone explodes (red curve), driving the growth of the total population (blue curve), while taking over a large fraction of the carrying capacity of the system, and then decays back as the infection is cleared.

On average, the effects of division and death almost balance each other, with a slight bias towards death because of the turnover imposed by thymic or bone marrow output. However, at a given time,
a clone that has high affinity for several present antigens will undergo a transient but rapid growth, while most other clones will decay slowly towards exctinction. In other words, locally in time, the antigenic environment creates a unique ``fitness'' for each clone. 
Since growth is exponential in time, these differential fitnesses can lead to very large differences in clone sizes, even if variability in antigen concentrations or affinities are nominally small. We thus expect to observe large tails in the distrubution of clone size. Fig.~\ref{secondgraph}A shows the cumulative probability distribution function (CDF) of clone sizes obtained at steady state (blue curve) showing a clear power-law behaviour for large clones, spanning several decades.

The exponent of the power-law is independent of the introduction size of clones (see inset of Fig.~\ref{secondgraph}A), and the specifics of the randomness in the environment (exponential decay, random number of partners, random interaction strength) as long as its first and second moment are kept fixed (See Fig.~\ref{micronoise} and Appendix \ref{microscopicnoise}).

\subsection{Simplified models and the origin of the power law}
To understand the power-law behavior observed in the simulations, and its robustness to various parameters and sources of stochasticity, we decompose the overall fitness of a clone at a given time (its instantaneous growth rate) into a constant, clone-independent part equal to its average $f_0<0$, and a clone-specific fluctuating part of zero mean, denoted by $f_i(t)$. This leads to rewriting Eq.~\ref{eq:simul} as:
\begin{equation}\label{eq:MF}
\frac{dC_{i}}{dt} =[f_0+f_i(t)]C_i(t)+B\xi_i(t),
\end{equation}
with $B^2\approx (|f_0|+2\mu)C_i$.

The function $f_i(t)$ encodes the fluctuations of the environment as experienced by clone $i$. Because antigens can be recognized by several receptors, these fluctuations may be correlated between clones. Assuming that these correlations are weak, $\langle f_i(t) f_j(t')\rangle\approx 0$, amounts to treating each clone independently of each other, and thus to reducing the problem to the single clone level.
The stochastic process giving rise to $f_i(t)$ is a sum of Poisson-distributed exponentially decaying spikes. This process is not easily amenable to analytical treatment, but we can replace it with a simpler stochastic process with the same temporal autocorrelation function.
This autocorrelation is given by $\langle f_i(t) f_i(t')\rangle = A^2 e^{- \lambda |t-t'|}$,
with the antigenic noise strength $A^2=s_{\rm A} p a_0^2 \< K^2\> \lambda^{-1}$, and where we recall that $\lambda^{-1}$ is the characteristic lifetime of antigens. The simplest process with the same autocorrelation function is given by an overdamped spring in a thermal bath, or Ornstein-Uhlenbeck process,
\beq\label{eq:OU}
\frac{df_i}{dt}=-\lambda f_i + \sqrt{2}\gamma \eta_i(t),
\eeq
with $\eta_i(t)$ a Gaussian white noise of intensity 1 and $\gamma=A\sqrt{\lambda}$ quantifies the strength of variability of the antigenic environment (see Appendix \ref{colored}).  This is also the process of maximum entropy or caliber \cite{Presse2013} with that autocorrelation function (see Appendix \ref{maxent} and \cite{AleksThierry2014}).

The effect of the birth death noise $B\xi_i(t)$ is negligible when compared to the fitness variations for large clones and it has no effect on the tail (see Fig.~\ref{bdnoeff} and Appendix \ref{whitenoise}). It can thus be ignored when looking at the tail of the distribution and its power law exponent, but it will play an important role for defining the range over which the power law is satisfied.

\begin{figure}
\noindent\includegraphics[width=\linewidth]{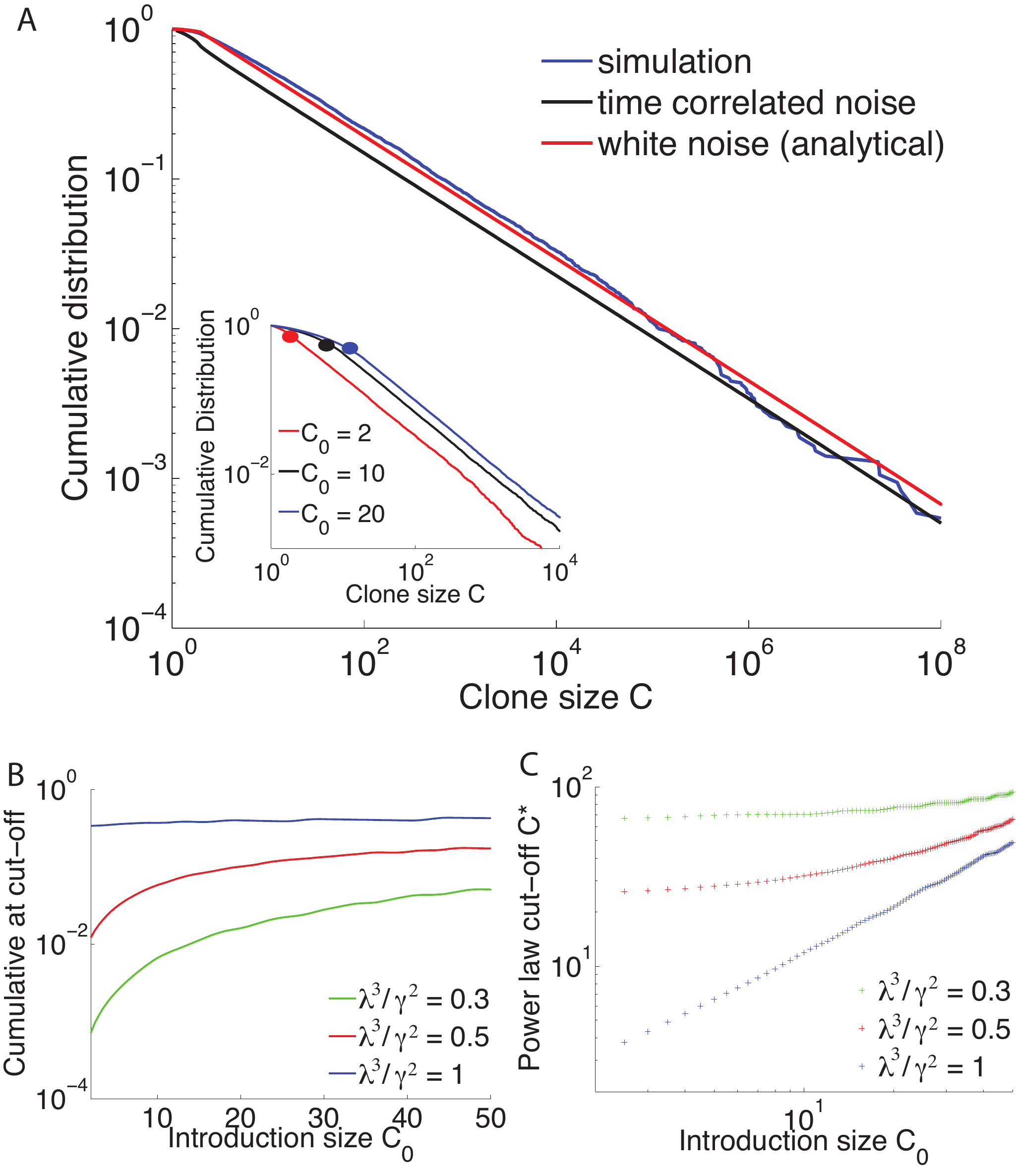}
\caption{Clone size distributions for populations with fluctuating antigenic, clone-specific fitness. {\bf A.} Comparison of simulations and simplified models of clone dynamics. Blue curve: cumulative distribution of clone sizes obtained from the simulation of Eq.~\ref{eq:simul}. Black curve: a simplified, numerically solvable model of random clone-specific growth, also predicts a power-law behaviour. 
Red curve: analytical solution fo the Gaussian white noise model, Eq.~\ref{eq:FP}. Parameters used: $\mu=0.98$ day$^{-1}$, $\nu=1.18$ day$^{-1}$, $\lambda=2$ day$^{-1}$,  $s_{\rm C}=2000$ day${}^{-1}$ , $C_0=2$, $s_{\rm A}=1.96\cdot 10^{7}$ day${}^{-1}$.  Inset: the exponent is independent of the initial clone size. Results from simulation with different values of the introduction clone size. The cut-off value of the power law behaviour, represented here as a dot, is strongly dependent on the value of $C_{0}$. Parameters are $\mu=0.2$ day$^{-1}$, $\nu=0.4$ day$^{-1}$, $\lambda=2$ day$^{-1}$, $\gamma=1$ day$^{-3/2}$ and $s_{\rm C}=5000$. {\bf B.} Value of the cumulative distribution function at the point of the power law cut-off as a function of the introduction clone size $C_{0}$ for different values of a dimensionless parameter related to the effective strength of antigen fluctuations relative to their characteristic lifetime $\lambda^3/\gamma^2$ for a fixed power law exponent $\alpha$. We use the cumulative distribution function because it is robust, invariant under multiplicative rescaling of the clone sizes. This way we do not need to correct directly for PCR multiplication or sampling. Parameters are for B and C $\mu=4.491$ days$^{-1}$, $\nu=5.489$ days$^{-1}$ and $\alpha=-0.998$. {\bf C.} Power-law cut-off as a function of the introduction clone size.
\label{secondgraph}}
\end{figure}

The population dynamics described by Eqs.~\ref{eq:MF} and \ref{eq:OU} can be reformulated in terms of a  Fokker-Planck equation for the joint abundance $\rho$ of clones of a given log-size $x=\log C$ and a given fitness $f$:
\begin{equation}\label{eq:FP}
\frac{\partial \rho(x,f,t)}{\partial t} = - (f_{0}+f) \frac{\partial \rho}{\partial x} + \lambda \frac{\partial (f \rho)}{\partial f} + \gamma^{2} \frac{\partial^{2} \rho}{\partial f^2} + s(x,f),
\end{equation}
where the source term $s(x,f)$ describes new clones arriving at rate $s_C$ with size $C_0$ and normally distributed fitnesses of variance $\<f^2\>=\gamma^2/\lambda$. This Fokker-Planck equation can be solved numerically with finite element methods with an absorbing boundary condition at $x=0$ to account for clone extinction. The solution, represented by the black curve in Fig.~\ref{secondgraph}A, matches closely that of the full simulated population dynamics (in blue). 
The power-law behaviour is apparent above a transition point that depends on the distribution of introduction sizes of new clones and the parameters of the model (see below). Intuitively, the microscopic details of the noise are not expected to matter when considering long time scales, as a consequence of the central limit theorem. However, the long tails of the distribution of clone sizes involve rare events and belong to the regime of large deviations, for which these microscopic details may be important. Therefore, the agreement between the process described by the overdamped spring and the exponentially decaying, Poisson distributed antigens is not guaranteed, and in fact does not hold in all parameter regimes (see Fig.~\ref{poigau}).

We can further simplify the properties of the noise by assuming that its autocorrelation time is small compared to other timescales. This leads to taking the limit $\gamma,\lambda\to \infty$ while keeping their ratio constant $\sigma=\gamma/\lambda$ constant, so that $f_i(t)$ is just a Gaussian white noise with $\<f_i(t)f_i(t')\>=2\sigma^2\delta(t-t')$ (see Appendix \ref{whitenoise} and Fig. \ref{limcolor}).
The corresponding Fokker-Planck equation now reads
\begin{equation}
\partial_{t} \rho(x,t) = - f_{0} \partial_{x} \rho(x,t)  +  \sigma^2 \partial_{x}^{2} \rho(x,t) + s(x),
\end{equation}
with $s(x)=s_C\delta(x-\log(C_0))$. 
This equation can be solved analytically at steady state, and the resulting clone size distribution is, for $C>C_0$:
\begin{equation}\label{powerlaw}
\rho(C) = \frac{s_C}{\alpha \sigma^2} \frac{1}{C^{\alpha +1}},
\end{equation}
with $\alpha = |f_{0}|/{\sigma^2}={\lambda|f_0|}/{A^2}$ (details in Appendix \ref{whitenoise}). The full solution, represented in Fig.~\ref{secondgraph}A in red, captures well the long-tail behaviour of the clone size distribution despite ignoring the temporal correlations of the noise, and approaches the solution of the colored-noise model (Eq.~\ref{eq:OU}) as $\lambda,\gamma\to\infty$, as expected (see Fig.~\ref{secondgraph}A).

The power law behaviour and its exponent depend on the noise intensity, but are otherwise insensitive to the precise details of the microscopic noise, including its temporal properties.
Fat tails (small $\alpha$) are expected when the average cell lifetime is long (small $|f_0|$) and when the antigenic noise is high (large $\sigma$ or $A$). The explicit expression for the exponent of the power law $1+\alpha$ as a function of the biological parameters can be used to infer the antigenic noise strength $A^2$ directly from data. An inverse lifetime of $|f_{0}|\approx 10^{-3}$ days$^{-1}$ of a typical T-cell can be estimated as the ratio of thymic output to the total population of lymphocytes in the body \cite{Goronzy2007,Westera2013,Vrisekoop2008}.
The characteristic lifetime of antigens $\lambda^{-1}$ is harder to estimate, as it corresponds to the turnover time of the antigens that the body is exposed to, but is probably of the order of days or a few weeks, $\lambda\approx 0.1$ day${}^{-1}$.
We estimated $\alpha=1\pm 0.2$ from the zebrafish data of Fig.~\ref{firstgraph}A \cite{quake-2009,mora2010} using canonical methods of power-law exponent extraction \cite{newmanclauset} (see Appendix \ref{dataanalysis} for details), and also found a similar value in human T cells \cite{Bolkhovskaya2014}. The resulting estimate, $A=10^{-2}$ day${}^{-1}$,
is rather striking, as it implies that fluctuations in the net clone growth rate, $A$, are much larger than its average $f_0$.

While the distribution always exhibits a power law for large clones, this behavior does not extend to clones of arbitrarily small sizes, where the details of the noise and how new clones are introduced matter. We define a power-law cut-off $C^*$ as the smallest clone size for which the cumulative distribution function (CDF) differs from its best power-law fit by less than $10\%$. Using numerical solutions to the Fokker-Planck equation associated to the colored-noise model, we can draw a map of $C^*$ as a function of the parameters of the system. In Fig.~\ref{secondgraph}B-C we show how $C^*$ varies as a function of the introduction size for different values of the dimensionless parameter related to the effective strength of antigen fluctuations relative to their characteristic lifetime at fixed power law exponents. In principle one can use this dependency to infer effective parameters from data. In practice, when dealing with data it is more convenient to consider the value of the cumulative distribution at $C^*$, rather than $C^*$ itself. For example, fixing $C_0=4$ and fitting the curve of Fig.~\ref{firstgraph}A with our simplified model using $\lambda$ as an adjustable parameter, we obtain $\lambda\approx 0.14$ day${}^{-1}$ (see Appendix~\ref{dataanalysis}), which corresponds to a characteristic lifetime of antigens of around a week. Although this estimate must be taken with care, because of possible PCR amplification biases plaguing the small clone size end of the distribution, the procedure described here can be applied generally to any future repertoire sequencing dataset for which reliable sequence counts are available.

\subsection{A model of fluctuating phenotypic fitness}
So far, we have assumed that fitness fluctuations are identical for all members of a same clone.
However, the division and death of lymphocytes do not only depend on signaling through their TCR or BCR. For example, cytokines are also growth inducers and homeostatic agents \cite{Schluns2000,Tan2001}, and the ability to bind to cytokines depends on single-cell properties such as the number of cytokine receptors on the membrane of a given cell, independent of their BCR or TCR receptor. Other stochastic single-cell factors may affect cell division and death. These signals and factors are {\em cell} specific, as opposed to the {\em clone} specific properties related to BCR or TCR binding. Together, they define a global phenotypic state of the cell that determines its time-varying ``fitness,'' independent of the clone and its T-cell or B-cell receptor.
This does not mean that these phenotypic fitness fluctuations are independent across the cells belonging to the same clone. Cells within a clone share a common ancestry, and may have inherited some phenotypic properties of their common ancestors, making their fitnesses effectively correlated with each other. However, this phenotypic memory gets lost over time, unlike fitness effects mediated by antigen-specific receptors.

We account for these phenotypic fitness fluctuations by a function $f_c(t)$ quantifying how much the fitness of an individual cell $c$ differs from the average fitness $f_0$. This fitness difference is assumed to be partially heritable, which we model by:
\begin{equation}\label{eq:OU2}
\frac{d f_c}{dt} = -\lambda  f_c(t) + \sqrt{2} \gamma_c \eta_c(t),
\end{equation}
where $\lambda^{-1}$ is the heritability, or the typical time over which the fitness-determining trait is inherited, $\gamma_c$ quantifies the variability of the fitness trait, and $\eta_c(t)$ is a cell-specific Gaussian white noise of power 1. Despite its formal equivalence with Eq.~\ref{eq:OU}, it is important to note that here the fitness dynamics occurs at the level of the single cell (and its offspring) instead of the entire clone. The dynamics of the fitness $f_i(t)$ of a given clone $i$ can be approximated from Eq.~\ref{eq:OU2} by averaging the fitnesses $f_c(t)$ of cells in that clone, yielding:
\begin{align}
\frac{dC_i}{dt} &= [f_{0}+f_i(t)] C_i (t) + \sqrt{(\nu+\mu)C_i(t)} \xi_i(t), \label{eq:IL1}\\
\frac{df_i}{dt} & = - \lambda f_i(t) + \frac{1}{\sqrt{C_i(t)}} \sqrt{2} \gamma_c \eta_i(t),\label{eq:IL2}
\end{align}
where $\eta_i(t)$ and $\xi_i(t)$ are clone-specific white noise of intensity 1, and $\nu$ and $\mu$ are the average birth and death rates, respectively, so that $f_0=\nu-\mu$ (details in Appendix \ref{cellspecifi_nomemory}). 
The difference with Eq.~\ref{eq:OU} is the
$1/\sqrt{C_i(t)}$ prefactor in the fitness noise $\eta_i(t)$, which stems from the averaging of that noise over all cells in the clone, by virtue of the law of large numbers. Because of this prefactor, the fitness noise is now of the same order of magnitude as the birth-death noise, which must now be fully taken into account. Taking Eq.~\ref{eq:IL1} and Eq.~\ref{eq:IL2} at the population level gives a Fokker-Planck equation with a source term accounting for the import of new clones. 
We verify the numerical steady state Fokker--Planck solution against Gillespie simulations (Fig \ref{il7example}, see Appendix \ref{cellspecificnoise} for details).

Fig.~\ref{thirdgraph}A-B show the distribution of clone sizes for different values of the phenotypic relaxation rate $\lambda$ and environment amplitude $\gamma_{c}$. These distributions vary from a sharp exponential drop in the case of low heritability (large $\lambda$) to heavier tails in the case of long conserved cell states (small $\lambda$). To quantify the extend to which these distributions can be described as heavy-tailed, we fit them to a power law with exponential cut-off, $\rho(C) \propto C^{-1-\alpha} e^{-C/C_m}$, where $C_m$ is the value below which the distribution could be interpreted as an (imperfect) power law. Fig.~\ref{thirdgraph}C shows a strong dependency of this cut-off with the phenotypic memory $\lambda^{-1}$. The longer the phenotypic memory $\lambda^{-1}$, the more clone-specific the fitness looks like, and the more the distribution can be mistaken for a power law in a finite-size experimental distribution. Larger birth-death noise also extends the range of validity of the power-law. As a result, and despite the absence of a true power-law behaviour, these models of  fluctuating phenotypic fitnesses cannot be discarded based on current experimental data.

\begin{figure}
\begin{center}
\noindent\includegraphics[width=\linewidth]{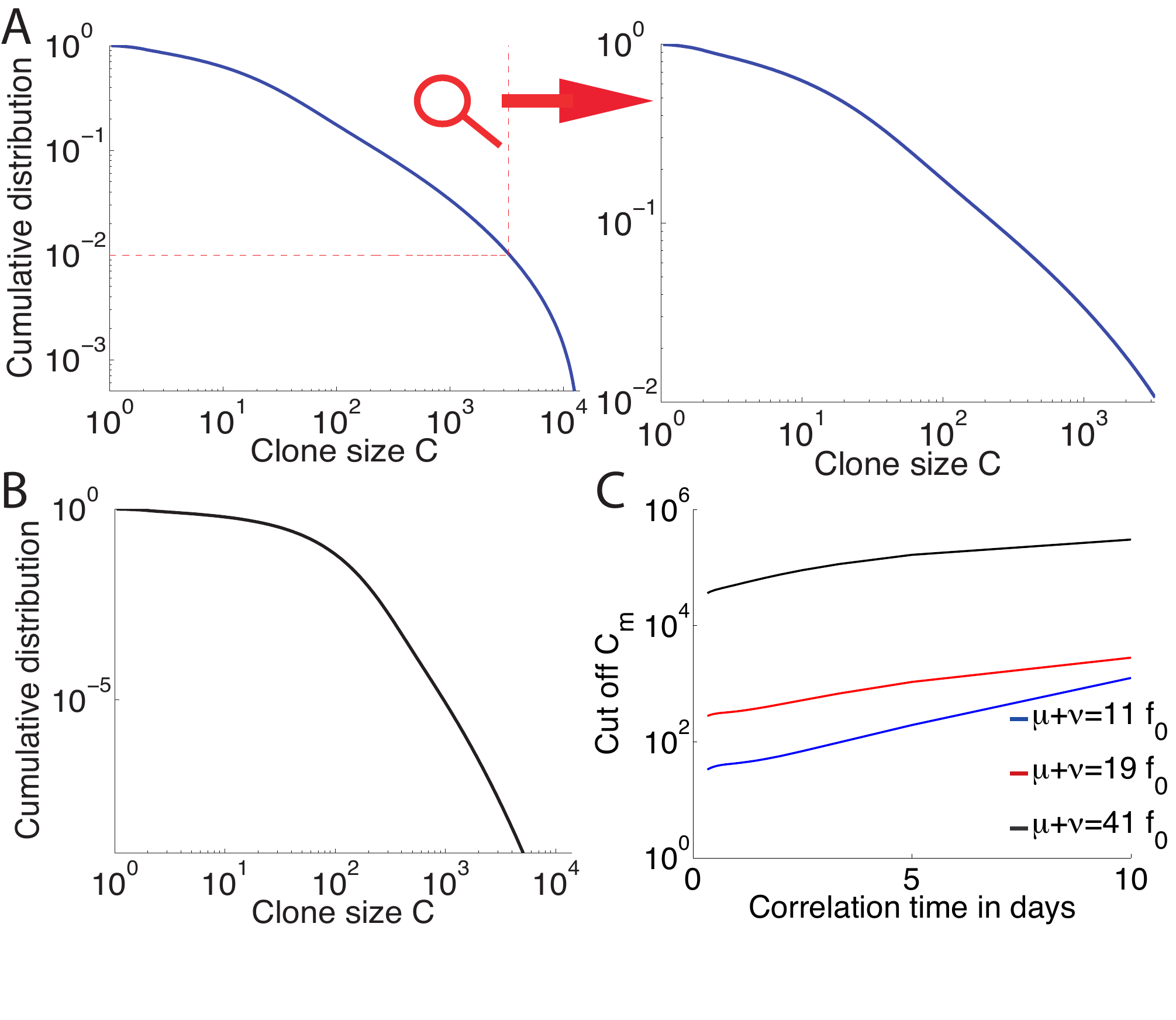}
\caption{Clone size distributions for populations with a cell specific fluctuating phenotypic fitness. {\bf A.} Cumulative distribution of clone sizes for moderate phenotypic heritability ($\lambda^{-1}$). The distribution is power-law like for small clone values and drops above a cut-off around $0.01$ of clone size probability. An experiment that does not sequence the repertoire deeply enough could report a power law behavior (see zoom). Parameters are $\mu=0.17$ days$^{-1}$, $\nu=0.3$ day$^{-1}$, $\lambda=0.4$ days$^{-1}$ and $\gamma_c=0.5$ days$^{-3/2}$. $C_0=2$ for all three graphs. 
{\bf B.} An example of a distribution of clone sizes from a cell-specific model with very low environmental noise, close to the pure birth-death limit. The distribution is flat ($\alpha=0$) and then drops exponentially. It does not resemble experimental data. Parameters are $\mu=0.1$ days$^{-1}$, $\nu=0.3$ days$^{-1}$, $\lambda=2$ days$^{-1}$ and $\gamma_c=5$ days$^{-3/2}$. 
{\bf C.} Value of the cumulative distribution at the exponential cut-off as a function of the speed of environment variations $\lambda$, for different birth-death noise levels. Parameters are $f_0=-0.998$ days$^{-1}$ and $f_0 \lambda^2/\gamma_c^2=0.998$. 
\label{thirdgraph}}
\end{center}
\end{figure}

The model can be solved exactly at the two extremes of the heritability parameter $\lambda$. In the limit of infinite heritability ($\lambda\to 0$) the system is governed by selective sweeps.
The clone with the largest fitness completely dominates the population, until it is replaced by a better one, giving rise to a trivial clone-size distribution. In the opposite limit, when heritability goes to 0 ($\lambda\to +\infty$), the Fokker-Planck equation can be solved analytically (see Appendices \ref{cellspecifi_nomemory} and \ref{purecellspe}), yielding an exact power-law with exponential cutoff, $\rho(C) \propto C^{-1-\alpha} e^{-C/C_m}$, with
$\alpha = -{[{1+ {(\mu+\nu) \lambda^2}/{2 \gamma_c^2}}]}^{-1}$
and
$C_m = (\mu - \nu)^{-1}[ {(\mu+d\nu)}/{2} + {\gamma_c^2}/{\lambda^2} ]$. The numerical solution of  Fig.~\ref{thirdgraph}B is close to this limit. Note that even with a negligible exponential cutoff, the predicted $\alpha<0$ contradicts experimental observations.

\section{Discussion}

Clone size distributions from high throughput data have been available for a few years and hold great promise for understanding the processes that shape the composition of the repertoire. Yet their interpretation requires models. The fate of lymphocytes making up immune repertoires is governed by a wide variety of mechanisms and pathways that vary across cell types and species. 
Previous population dynamics approaches to repertoire evolution have taken great care in precisely modeling these processes for each compartment of the population, through the various mechanisms by which cells grow, die and change phenotype \cite{Stirk2010,Stirk2008, perelson-2001}. However, one of the most striking properties of repertoire statistics revealed by high-throughput sequencing---the observation of power laws in clone size distribution---hold true for various species (human, mice, zebrafish), cell type (B and T cells) and subsets (naive and memory, CD4 and CD8), and seems to be insensitive to these context-dependent details. These observations call for a stochastic description of the various 
properties affecting cell fate, and their ubiquity points to some universal features of the dynamics.

The model introduced in this paper describes the stochastic nature of the immune dynamics with a minimal number of parameters, easing the numerical and analytical treatment and helping interpret the different regimes. These parameters are effective in the sense that they integrate different levels of signaling, pathways, and mechanisms. We assumed that they are general enough that different cell types (B and T cells) or subsets can be described by the same dynamical equations.
Most of these parameters have natural interpretations: the typical lifespan of an infection or antigen stimulation is given by $\lambda^{-1}$, and the effect of these stimulations by $A$. For memory cells, $\lambda^{-1}$ corresponds to the duration of an infection. For naive cells, the interpretation is less clear, as non-pathogenic antigens such as self-antigens may not be as transient as pathogens; in this case $\lambda^{-1}$ could correspond to the characteristic timescale of changes in a clone's micro-environment. Likewise, in the context of cell-specific noise, the parameters $\lambda^{-1}$ and $\gamma_c$ correspond to the correlation time and variability of the cell phenotype. All these parameters may vary greatly across cell types and species.

We showed that the relevant sources of stochasticity for the shape of the clone--size distributions  fall into two main categories, depending on how cell fate is affected by the environment. 
 Either the stochastic elements of clone growth act in a clone-specific way, through their receptor (BCR or TCR), leading to power-law distributions with exponent $\geq 1$, or in a cell-specific way, {\em e.g.} through their variable level of sensitivity to cytokines (and more generally through any phenotypic trait affecting cell fitness), leading to exponentially decaying distributions with a power-law prefactor. These two types of signals (clone specific and cell specific) are important for the somatic evolution of the immune system \cite{Schluns2000,Tan2001, Seddon2002, Tanchot1997, Nesic1998, Freitas1995}
and our analysis shows that the shape of the clone size distribution is informative of their relative importance to  the repertoire dynamics. It provides a first theoretical setting and an initial systematic classification for modeling immune repertoire dynamics. Our method applied to high-throughput sequencing data can be used
to quantify how much each type of signal contributes to the overall dynamics, and what is the driving force for the different cell subsets.
 For example, although it is reasonable to speculate that clone-specific signals should dominate for memory cells (through antigen recognition), and cell-specific selection for naive cells (through cytokine-mediated homeostatic division), the relative importance of these signals for both cell types is yet to be precisely quantified, and may vary across species.
A clear power law over several decades would strongly hint at dynamics dominated by interactions with antigens, while a faster decaying distribution would favor a scenario where individual cell fitness fluctuations dominate. Applying these methods to data from memory cells can give orders of magnitude for the division and half-life of memory lymphocytes, as well as the typical number of cells $C_0$ from a clone that are stored as memory following an infection.

The application of our method to data from the first immune repertoire survey (B cell receptors in zebrafish \cite{quake-2009}) suggests that clone-specific noise dominates in that case, allowing us to infer a relation between the dynamical parameters of the model from the observed power-law exponent $\approx 2$.  However, there are a few issues with applying our method directly to data in the current state of the experiments.
First, the counts ({\em i.e.} how many cells have the same receptor sequence and belong to the same clone) from many high-throughput repertoire sequencing experiments are imperfect because of PCR bias and sampling problems. New methods using single-molecule barcoding have been developed for RNA sequencing  \cite{Best2014,Vollmers2013,mamedov-2014}, but they do not solve the problem entirely, as the number of expressed mRNA molecules may not faithfully represent the cell numbers because of possible expression bias. In addition, most studies (with the exception of \cite{Dekosky2014}) have been sequencing only one of the two chains of lymphocyte receptors, which is insufficient to determine clone identity unambiguously. As methods improve, however, our model can be applied to future data to distinguish different sources of fitness stochasticitiy and to put reliable constraints on biological parameters.

Thanks to its generality, our model is also relevant beyond its immunological context, and follows previous attempts to explain power laws in other fields \cite{Sornette1997,Marsili1998,Mitzenmacher2004}.
The dynamics described here corresponds to a generalization of the neutral model of population genetics \cite{kimurabook} where thymic or bone marrow outputs are now reintepreted as new mutations or speciations, and where we have added a genotypic or phenotypic fitness noise (receptor or cell-specific noise, respectively). It was recently shown that such genotypic fitness noise strongly affects the fixation probability and time in a population of two alleles \cite{Cvijovi2015, Melbinger2015}. Our main result (Eq.~\ref{powerlaw}) shows how fitness noise can cause the clone-size distribution (called frequency spectrum in the context of population genetics) to follow a power law with an {\em arbitrary} exponent $>1$ in a population of fixed size, while the classical neutral model gives a power law of exponent 1 with an exponential cut off (as shown in our exact solution with $\gamma_c=0$). Our results can be used to explain complex allele frequency spectra using fluctuating fitness landscapes.

{\bf Acknowledgements.} This work was supported in part by grant ERCStG n. 306312.

\appendix

\setcounter{figure}{0}
\makeatletter 
\renewcommand{\thefigure}{S\@arabic\c@figure}
\makeatother

\section{Simple birth-death process with no fitness fluctuations, and its continuous limit}
\label{appbirthdeath}
In this Appendix we derive the steady-state clone size distribution for a system that does not experience any environmental stimulation or noise, but is governed by a birth death process. We will show that the small number fluctuations arising from the discrete nature of birth and death are not sufficient to explain the observed distributions. We also  show that our choice of a continuous birth death process is equivalent to its discrete version. 

  \begin{figure}
\includegraphics[width=0.5\textwidth]{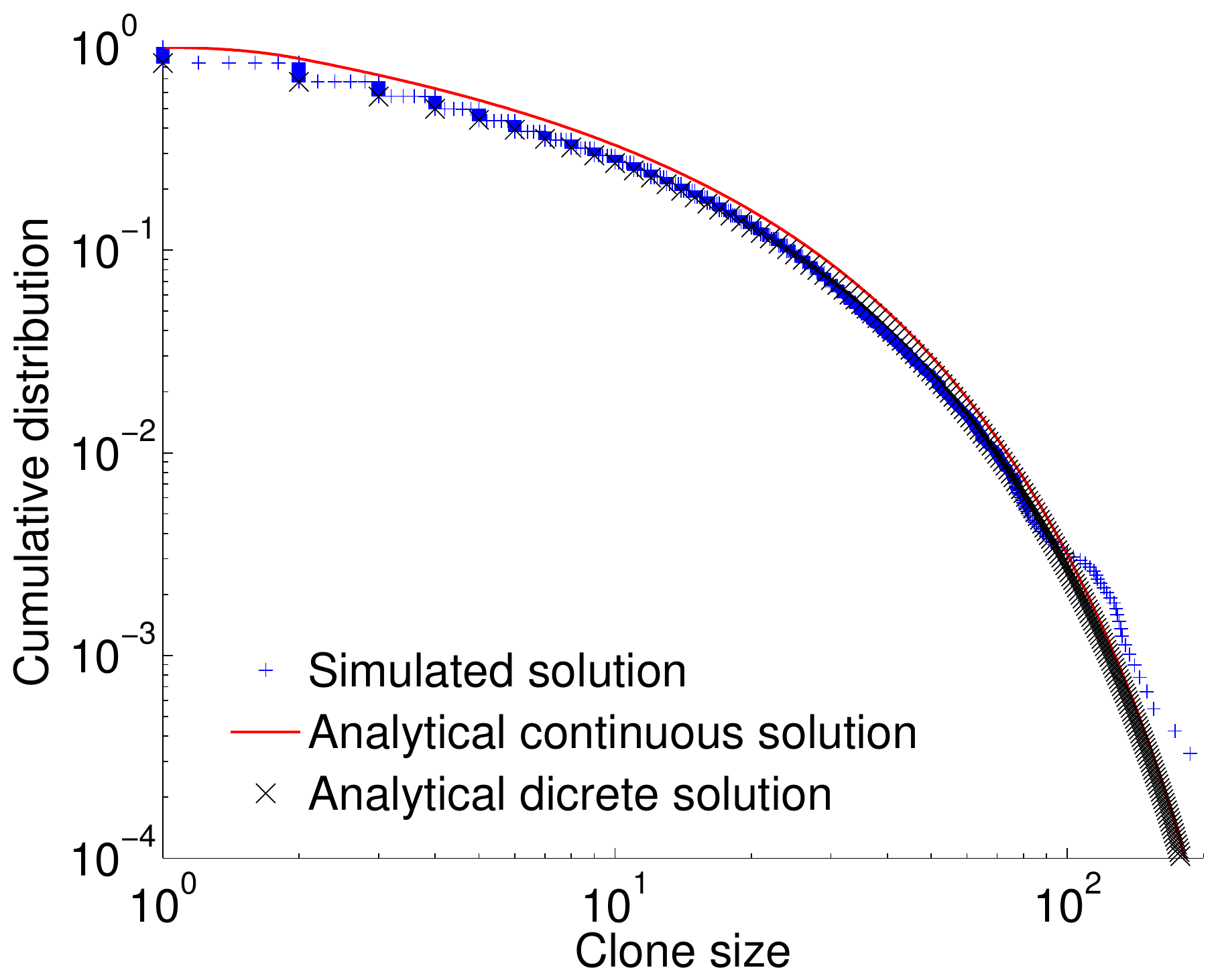}
\caption{We compare results from a full Gillespie simulation (blue crosses) of a system with only birth-death dynamics with analytical prediction for a discrete system (black crosses, Eq.~\ref{discretebd}) and a continuous system (red curve, Eq.~\ref{continuousbd}). The prediction with discrete variables is more accurate for small clones but the behaviour of all systems is the same for large populations. The parameters are $\mu = 1.45$ day$^{-1}$, $\nu=1.5$ day$^{-1}$, $C_0 = 2$ and we introduce $2000$ new clones per day.  \label{birthdeathgraph}}
\end{figure}

The multiplicative birth--death process corresponds to the following discrete dynamics: 
\begin{equation}
  \begin{cases}
     P( n \to n+1 ) = \mu n dt  \\ 
   P(n \to n-1) = \nu n dt,
 \end{cases}
\end{equation}
where $\mu$ is the division rate, $\nu$ the death rate. We assume that the population of cells of size $n$ is maintained out of equilibrium by a source of new cells. 
The steady state solution for cell numbers above the value of the source satisfies detailed balance \begin{equation}
P(n) \mu n = P(n+1) \nu (n+1)
\end{equation}
and, assuming  the death rate is larger than the birth rate, takes the form
\begin{equation}
\label{discretebd}
P(n) \sim \frac{K}{n} e^{ -n \log{\nu/\mu}  }.
\end{equation}

The continuous counterpart of this discrete stochastic process corresponds to the following linear-noise approximation:
\begin{equation}
\partial_{t} C_i = f_{0} C_i + \sqrt{(\mu+\nu)C_i} \xi,
\end{equation}
where $\<\xi_i(t)\xi_i(t')\>=\delta(t-t')$ and $f_{0} = \mu-\nu <0$ (and we use the \^Ito convention ). 
In terms of $x= \log C$ the Langevin equation is
\begin{equation}
\partial_{t} x = f_{0} + \sqrt{\mu+\nu} e^{-x/2} \xi - e^{-x} \frac{(\mu+\nu)}{2},
\end{equation}
and the corresponding Fokker-Planck equation reads
\begin{equation}
\partial_{t} \rho = \partial_{x} ( - f_{0} \rho  ) + \partial_{x}^{2} \left( \frac{\mu+\nu}{2} e^{-x} \rho     \right) + \partial_{x} \left( e^{-x} \rho \frac{\mu+\nu}{2} \right) + s(x),
\end{equation}
where $s(x)$ is the distribution of sizes of newly arriving clones. At steady state, we find
\begin{equation}
K - s_{C} \theta(x- x_{0}) = - f_{0} \rho  + \frac{\mu+\nu}{2}   e^{-x} \rho',
\end{equation}
where $K$ is an integration constant. Defining
\begin{equation}
C_m = ({\mu+\nu})/({2 |f_0|})
\end{equation}
 for $x<x_{0}$ we obtain
\begin{equation}
\rho(x) = e^ {- e^{x}/C_m} K \int_{0}^{x} e^{x} e^{ e^{x}/C_m } = K C_m ( 1 - e^{- (e^{x}-1)/C_m}  )
\end{equation}
and for $x>x_{0}$ 
\begin{eqnarray}
\rho(x) &= &e^{ -  e^{x} /C_m } C_m \Big[ K e^{ e^{x} / C_m}  - Ke^{1/C_m} \\ \nonumber
&&-  \frac{ s_{C}}{|f_{0}|C_m} e^{ e^{x}/C_m} +  \frac{ s_{C}}{|f_{0}| C_m} e^{ e^{x_{0}}/C_m}  \Big] 
\end{eqnarray}
To ensure convergence we set $K ={ s_{C}}/({|f_{0}|C_m}) $ and the steady solution of the Fokker-Planck equation is
\begin{equation}
    \rho(x)  = 
    \begin{cases}
   \frac{s_{C}}{|f_{0}|} (1- e^{ - (e^{x}-1) /C_m })  \text{ ,   if     } x<x_{0}  \\ 
    \frac{s_{C}}{|f_{0}|} (e^{  e^{x_{0}} /C_m }- e^{C_m^{-1}})  e^{-  e^{x} /C_m}  \text{,   if   } x>x_{0}
    \end{cases}
\end{equation}
or in terms of the clone size
\begin{equation}
\label{continuousbd}
    \rho(C)  = 
    \begin{cases}
  \frac{1}{ C} (1- e^{ - (C-1)/C_m  }) \text{,     if   } C<C_{0}  \\ 
  (e^{  C_{0} /C_m }- e^{C_m^{-1}})  \frac{e^{-  C /C_m}}{C}   \text{, if   } C>C_{0}
    \end{cases}
\end{equation}
This result is exactly equivalent to that of Eq.~\ref{discretebd} when $\nu-\mu=|f_0|\ll \mu,\nu$.
The accuracy of the approximation is verified in Fig. \ref{birthdeathgraph}.
Even for very large  exponential cutoff values, $C_{m}$, the apparent exponent is $\alpha = 0$, corresponding to a flat cumulative distribution. This distribution  is inconsistent with experiments, regardless of sequencing depth and we conclude that pure birth-death noise is not sufficient to explain the observed distributions.

\section{Effects of explicit global homeostasis}
\label{homeostasis}

\begin{figure}
\includegraphics[width=0.5\textwidth]{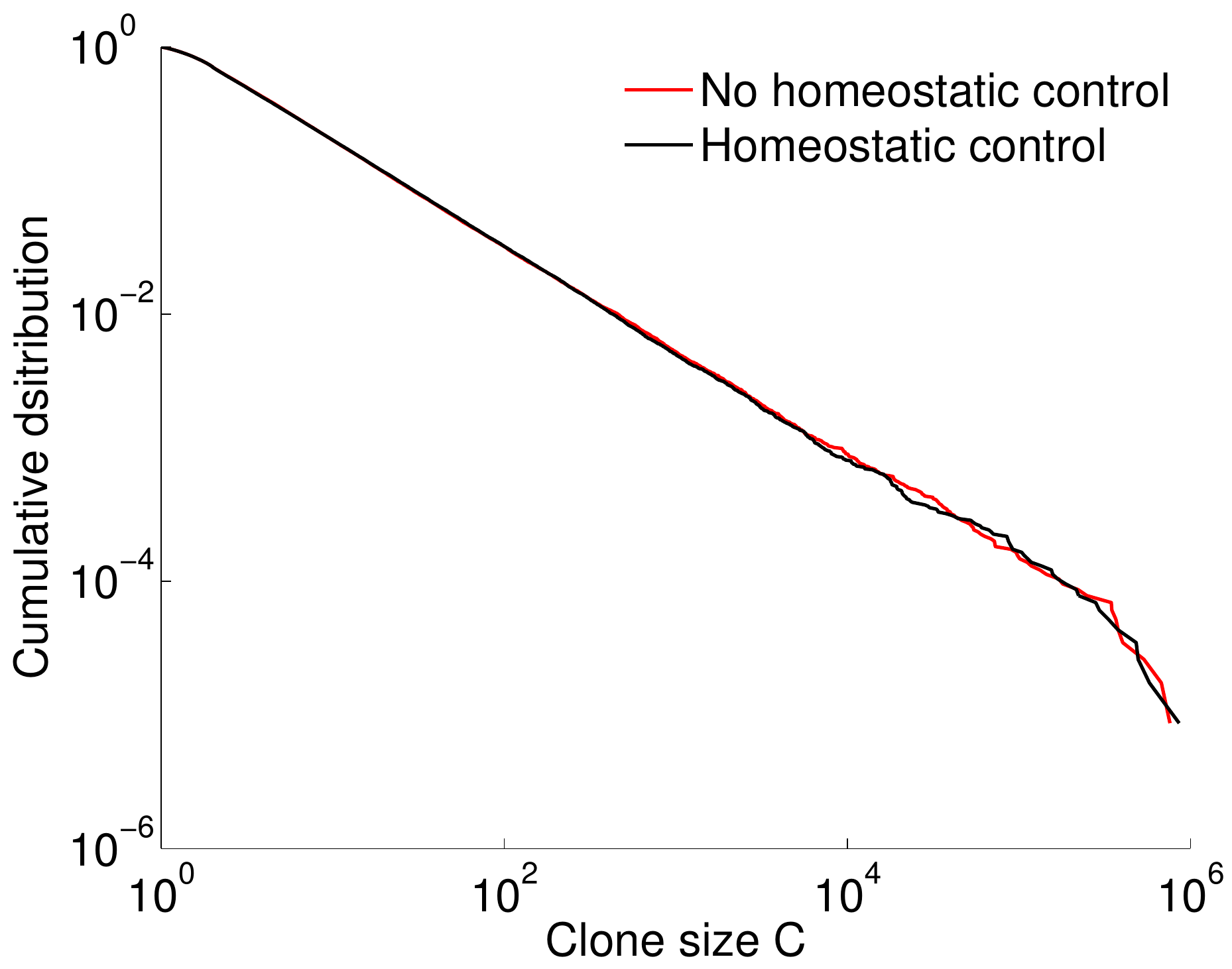}
\caption{Adding an explicit homeostatic control term does not affect the clone size distribution compared to tuning the degradation and death rates to obtain a given repertoire size as is done in the main text. Comparison of the clone size distribution with an explicit homeostatic control term given by Eq.~\ref{homeoterm} (black line) to the distribution presented in the main text (red line). We simulate the Langevin equation for a division rate $\mu = 0.2$ days$^{-1}$, death rate $\nu=0.4$ days$^{-1}$, introduction size $C_0 = 2$, environmental correlation time of $\lambda^{-1}=0.5$ days and an amplitude of variations of the environment $A = 1.41 $ days$^{-1}$ without any homeostatic control for the red curve and with carrying capacity $N=4 \cdot10^{10}$ ($h=1$) and a homeostatic exponent $r=3$ for the black curve.}
\label{homeocon}
\end{figure}

In the simulations of clone dynamics in a fluctuating environment presented in the ``Clone dynamics in a fluctuating antigenic landscape'' Results section of the main text, we did not explicitly include a homeostatic control term, but tuned the division and death rates to achieve a given repertoire size. Here we add  an explicit homeostatic term to the growth and degradation terms in the Langevin simulations described by Eq.~1 of the main text
\begin{equation}
- h \left[ \frac{\sum_i C_{i}}{N} \right]^r,
\label{homeoterm}
\end{equation}
where $N$ is a carrying capacity, $h$ is the homeostatic constant multiplicator and $r$ is the exponent of homeostatic response that described the sharpness of the response when approaching then carrying capacity limit. Comparing in Fig. \ref{homeocon} the resulting clone size distribution obtained with the explicit homeostatic term to the distribution from the simulations in the main text, we see that the explicit homeostatic term does not have an effect on the form of the distribution. It does have an effect on the trajectory of certain clones, and in particular on the response of the system to a very large invasion, making it an important feature of the dynamics of the immune system. However, as shown by the results in  Fig.~\ref{homeocon} its net effect on the clone size distribution can be taken into account by tuning   division and death. When considering specific trajectories in the mean field approximation homeostatic control will add a systematic negative drift to the clonal population and can be accounted for by an additional contribution to $ f_{0} $.

\section{Details of noise partition do not influence the clone size distribution function}
\label{microscopicnoise}
\begin{figure}
\includegraphics[width=0.5\textwidth]{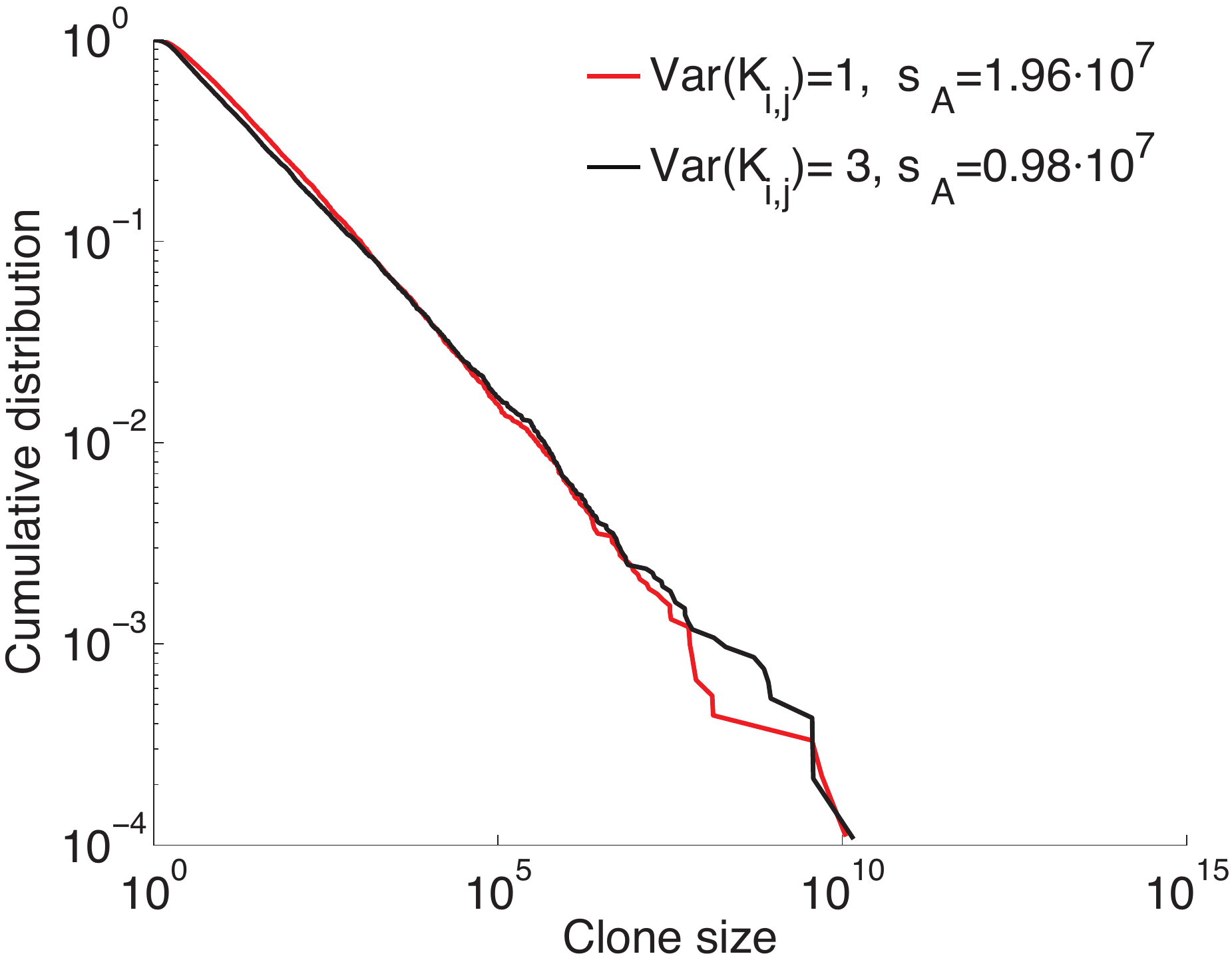}
\caption{Repartitioning the sources of stochasticity between the number of new antigens per time unit or the variability of binding probabilities does not influence the clone size distributions. We compare simulations of the full system dynamics defined by Eq.~1 of the main text with two sets of values $s_{\rm A}$ of  the poisson distributed number of newly arriving antigen $N_a$ and the variance of the Gaussian distributed binding probabilities $K_{ij}$ that give the same total environmental noise $A^2=s_{\rm A} p a_0^2 \< K^2\> \lambda^{-1}$. The parameters were taken to be (as in Fig.~1) $s_{\rm C}=2000$ day${}^{-1}$ , $C_0=2$, day${}^{-1}$, $a_{j,0}=a_0=1$, $\lambda=2$  day${}^{-1}$, $p=10^{-7}$, $\nu=0.98$ day${}^{-1}$, $\mu=1.18$ day${}^{-1}$. For the red curve the variance of the entries of $K_{ij}$ is $1$, so that $\< K^2\>=2$ and $s_{\rm A}=1.96\cdot 10^{7}$ while for the black curve the variance of the entries of $K_{ij}$ is $3$, so that $\< K^2\>=4$, and $s_{\rm A}=0.98\cdot 10^{7}$.
 \label{micronoise}}
\end{figure}

In the simulation of the dynamics of receptors experiencing a clone-specific fitness presented in the ``Clone dynamics in a fluctuating antigenic landscape'' Results section of the main text  we distributed the noise between the different random distributions: the poisson distributed number of new antigens ($s_{\rm A}$), the variance of the initial concentrations ($a_{j,0}$) and the variance of the binding probability (the values of $K_{ij}$). We made specific choices for this reparation by picking specific parameters of the random processes.  Here we show that these specific choices of repartitioning the contributions to the noise do not influence the clone size distributions. Fig.~\ref{micronoise} compares clone size distributions obtained with different values of the poisson distributed number of newly arriving antigen $N_a$ and the variance of the Gaussian distributed binding probabilities $K_{ij}$, reproducing the same distributions in both cases.

\section{Model of temporally correlated clone-specific fitness fluctuations}
\label{colored}
In the ``Simplified models and the origin of the power law'' Results section of the main text we make a series of approximations to effectively describe the dynamics of immune cells: we first approximate the antigenic environment by a random process with time correlated (colored) noise and we later neglect these temporal correlations. In this section and Appendix \ref{whitenoise} we give the details that lead to the specific forms of the effective equations. In this Appendix we derive the Fokker-Planck equations for the time correlated noise model. In Appendix \ref{whitenoise} we will consider the limit of an infinitely quickly changing environment.

The Langevin equations describing the dynamics of cells experiencing clone specific fitness fluctuations with a finite correlation time are
\begin{align}
\frac{dC_i}{dt} &= [f_{0}+f_i(t)] C_i (t) + \sqrt{(\nu+\mu)C_i(t)} \xi_i(t), \label{eq:SI2d1}\\
\frac{df_i}{dt} & = - \lambda f_i(t) + \sqrt{2} \gamma \eta_i(t),\label{eq:SI2d2}
\end{align}
where $\<\xi_i(t)\xi_i(t')\>=\delta(t-t')$ represents birth death noise in the linear-noise approximation (with the \^Ito convention) and $\<\eta_i(t)\eta_i(t')\>=\delta(t-t')$ is the noise of antigenic environment. 
The autocorrelation function of this Ornstein-Uhlenbeck process is
\begin{equation}
\langle f_i(t)f_i(t') \rangle = e^{- \lambda (t+t')} \left( \langle f_i(0)^{2} \rangle -\frac{\gamma^{2}}{\lambda} \right) +  \frac{\gamma^{2}}{\lambda} e^{-\lambda |t-t'|} .
\label{OUp}
\end{equation}
We  pick the steady-state value of the  initial fitness distribution to cancel the first in Eq.~\ref{OUp}$, \langle f_i(0)^{2} \rangle = {\gamma^{2}}/{\lambda}$ and obtain
\begin{equation}
\langle f_i(t)f_i(t') \rangle  =  \frac{\gamma^{2}}{\lambda} e^{-\lambda |t-t'|},
\end{equation}
(conditioned on the integral of the net growth rate $f+f_{0}$ being positive so that the clone does not go extinct).
Setting $x=\log C$, we obtain  a new set of Langevin equations
\begin{align}
\partial_{t} x_{i} & = f_{0} + f_i + \sqrt{\mu+\nu} e^{-x_{i}/2} \xi_{i} - e^{-x_{i}} \frac{(\mu+\nu)}{2} , \\
\frac{df_i}{dt} & = - \lambda f_i + \sqrt{2} \gamma \eta_i,
\end{align}
where the birth-death noise is now treated in the  \^Ito convention.
The corresponding Fokker-Planck equation for the distribution of fitness and clone size at time $t$, $\rho(x,f,t)$, verifies
\begin{eqnarray}
\partial_{t} \rho&=& \partial_{x} (- f_{0} \rho) + \partial_{f} ( \lambda f \rho ) +  \partial_{f}^2 (\gamma^2 \rho) + \\ \nonumber
&&  \partial_{x}^{2} \left( \frac{\mu+\nu}{2} e^{-x} \rho     \right)   + \partial_{x} \left( e^{-x} \rho \frac{\mu+\nu}{2} \right) \\ \nonumber
&&+ s(x,f),
\end{eqnarray}
where $s(x,f)$ is the source of new clones. We solve this equation numerically using finite element methods to obtain clone size distributions for the clone-specific fitness model.

\section{The Ornstein Uhlenbeck process and maximum entropy}
\label{maxent}

In this Appendix we show that the maximum entropy or maximum caliber process with autocorrelation function $\langle x(t) x(t+s) \rangle =  A^2 e^{-\lambda |s|}$ corresponds to the Ornstein-Uhlenbeck process. We consider this continuous maximum entropy process as the continuous limit of a simpler maximum entropy system in discrete time. Burg's maximum entropy theorem \cite{infotheobook} states that the maximum entropy process in discrete time that constrains 
 $\langle X_n (t)^2 \rangle =  A^2 $ and $\langle X_n (t) X_{n+1} (t) \rangle =  A^2 e^{-\lambda \tau }$ corresponds to the following Markovian dynamics:
\begin{equation}\label{OUdiscrete}
X_{n+1} = e^{-\lambda \tau} X_n + \sqrt{1- e^{- 2 \lambda \tau}} A \eta,
\end{equation}
where $\eta$ is Gaussian white noise. In the limit of $\tau \rightarrow 0$ we recover the constrained autocorrelation function in the vicinity of $s=0^+$: $\<x(t)^2\>=A^2$, $(d/ds)\<x(t)x(t+s)\>|_{s=0^+}=-\lambda A^2$, and Eq.~\ref{OUdiscrete} converges to an Ornstein-Uhlenbeck process.

\section{Model solution for white-noise clone-specific fitness fluctuations}
\label{whitenoise}
In the limit of infinitely quickly fluctuating environments, $\gamma \rightarrow + \infty$ and $\lambda \rightarrow + \infty$ while keeping their ratio $\sigma= \gamma/\lambda$ constant, the autocorrelation of the fitness noise approaches a Dirac delta function, and the fluctuating part of the growth rate $f_i(t)$ converges to Gaussian white noise, $\<f_i(t)f_i(t')\>=2\sigma^2\delta(t-t')$. Effectively the immune cell dynamics are now described by 
a one dimensional Langevin equation for the clone size
\begin{equation}\label{eq:SI1d}
\partial_{t} C_{i} = f_{0} C_{i} + \sqrt{2} \sigma C_{i}  \eta_{i} + \sqrt{(\nu+\mu)C_i(t)} \xi_i,
\end{equation}
where $\<\eta_{i}(t)\eta_i(t')\>=\delta(t-t')$ follows the  Stratanovich convention and $\xi_i$ is as before.
The equation for the logarithm of the clone size $x=\log C$ is
\begin{equation}
\partial_{t} x_{i} = f_{0} + \sqrt{2} \sigma  \eta_{i} + \sqrt{\mu+\nu} e^{-x_{i}/2} \xi_i - e^{-x_{i}} \frac{(\mu+\nu)}{2}.
\end{equation}

\begin{figure}
\includegraphics[width=0.5\textwidth]{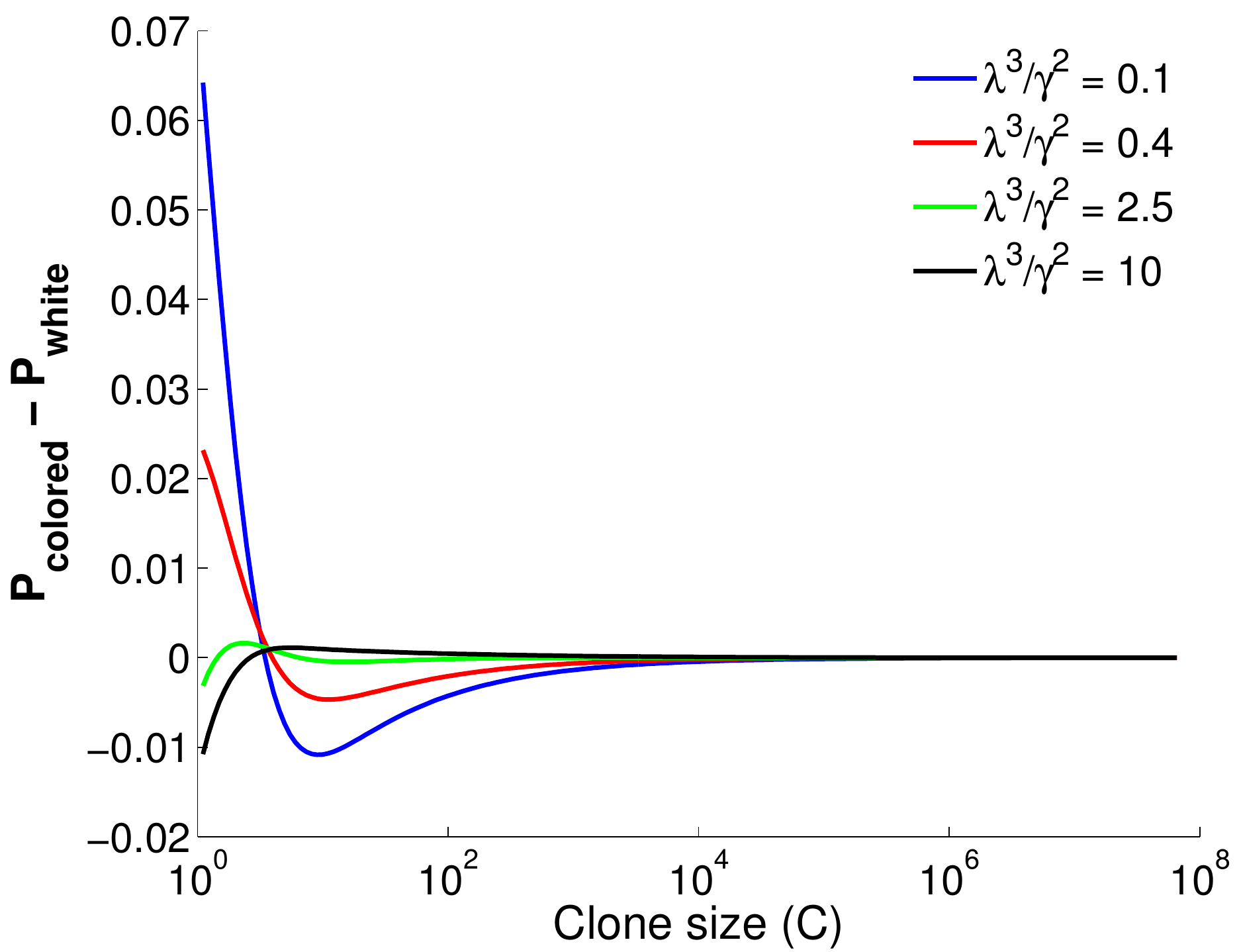}
\caption{Comparison between clone size distribution obtained as solutions of the time-correlated and time-uncorrelated noise models (without birth death noise). As the values of the dimensionless parameter related to the effective strength of antigen fluctuations relative to their characteristic lifetime $\lambda^3/\gamma^2$ grow the time correlated noise prediction converges to the exact power-law solution of the white-noise model. The cut-off value of the power law decreases with  $\lambda^3/\gamma^2$. All simulations performed at a constant value of $\alpha=|f_{0}| \lambda^2 / \gamma^2$ set to $0.5$. The value of $f_0$ is kept fixed to  $- 0.5$ days$^{-1}$ for all solutions. \label{limcolor}}
\end{figure}

We explicitly checked that the numerical solution to the clone specific fitness model in  Eqs.~\ref{eq:SI2d1} and \ref{eq:SI2d2} converged to the dynamics described by Eq.~\ref{eq:SI1d}, as demonstrated in Fig.~\ref{limcolor}. 

We now solve this equation analytically, starting with the case of no birth-death noise: 
Eq.~\ref{eq:SI1d} simplifies to
\begin{equation}
\partial_{t} C_{i} = f_{0} C_{i} + \sqrt{2} \sigma C_{i}  \eta_{i} 
\end{equation}
The equation for $x = \log{C}$ (using the Stratanovich convention) is
\begin{equation}
\partial_{t} x_i = f_{0} + \sqrt{2} \sigma \eta_i,
\end{equation}
with the corresponding Fokker Planck equation 
\begin{equation}
\partial_{t} \rho (x,t) = \partial_{x} (- f_{0} \rho) + \frac{1}{2} \partial_{x} [ 2 \sigma^2 \partial_{x} \rho  ] + s(x),
\end{equation}
where $s(x)$ is the source term describing the size of newly introduced clones. 
Assuming a constant initial clone size, $s(x)=s_C\delta(x-x_0)$, the steady state solution is
\begin{equation}
\rho(x) = e^{-\alpha x} \frac{1}{\alpha} \left[   K e^{\alpha x} - K - s_{C} \sigma^2 e^{\alpha x} + s_{C} \sigma^2 e^{x_{0}}   \right],
\end{equation}
where we have defined
\begin{equation}
\alpha = |f_{0}|  / \sigma^2,
\end{equation}
and $K$ is an  integration constant.
Imposing that $\rho$ vanishes at infinity sets $K = s_{C} \sigma^2 $ and the final form of the steady state clone size distribution is
\begin{eqnarray}
  \rho(x) & =& 
  \begin{cases}
  \frac{s_{C}}{|f_0|} \left(1 - e^{-\alpha x} \right)  \text{   if   } x<x_{0}  \\ 
 \frac{s_{C}}{|f_0|} e^{-\alpha x } \left(e^{x_{0}} - 1 \right) \text{   if   } x>x_{0},
 \end{cases}
\end{eqnarray}
or in terms of clone size $C=e^x$,
\begin{eqnarray}
   \rho(C) & =& 
  \begin{cases}
   \frac{s_{C}}{|f_0|C} \left(1 - \frac{1}{C^{\alpha}} \right)  \text{   if   } C<C_{0}  \\ 
 \frac{s_{C}}{|f_0|} \frac{1}{C^{\alpha+1}} \left( \frac{1}{C^{x_{0}}} - 1 \right) \text{   if   } C>C_{0}.
 \end{cases}
\end{eqnarray}

In all simulations and solutions we find that for large clones, the model of temporally correlated fitness fluctuations behaves as the its white noise limit. This behaviour can be explained by the 
fact that large clones need a long time to become large. At these long timescales, the characteristic time of noise correlation is negligible and the noise may be approximated as white. For this reason, the exponent $\alpha$ of the power law computed assuming a white noise for the fitness fluctuations is still valid even when that noise is actually correlated in time.

\begin{figure}
\includegraphics[width=0.5\textwidth]{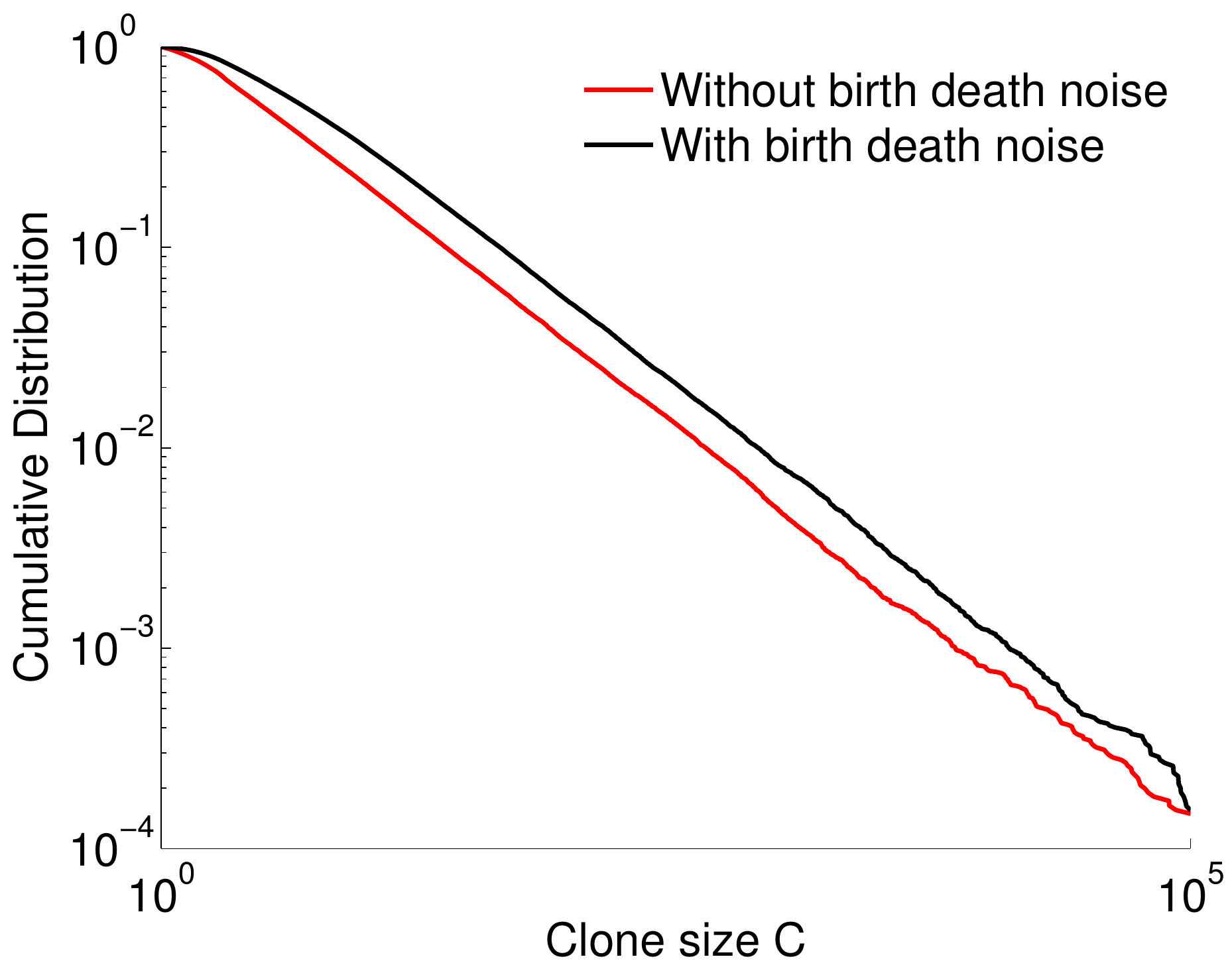}
\caption{We compare simulations of the Langevin dynamics with time correlated antigenic noise with birth-death noise (black line) to the same dynamics without the birth-death noise (red line). All other parameters are kept fixed.We find similar values of the power law exponents  but different small clone behaviours. The parameters are            
$\mu = 0.2$ day$^{-1}$, $\nu=0.4$ day$^{-1}$ (for red curve simply $f_0 = -0.2$ day$^{-1}$) , $C_0 = 2$, $\lambda = 2$ day$^{-1}$ and $\gamma=1$ day$^{-3/2}$
  \label{bdnoeff}} 
\end{figure}

Next, we re-introduce the birth-death noise and solve the general equation. The Langevin equation for $x=\log C$,
\begin{equation}
\partial_{t} x = f_{0} + \sqrt{2} \sigma \eta + \sqrt{\mu+\nu} e^{-x/2} \xi - e^{-x} \frac{(\mu+\nu)}{2} 
\end{equation}
 results in the Fokker-Planck equation for the distribution of clone sizes
\begin{equation}
\begin{split}
\partial_t  \rho= \partial_{x} (- f_{0} \rho) + \frac{1}{2} \partial_{x} [ 2 \sigma^2 \partial_{x} \rho  ] +  \partial_{x}^{2} \left( \frac{\mu + \nu}{2} e^{-x} \rho     \right)  \\ + \partial_{x} \left( e^{-x} \rho \frac{\mu + \nu}{2} \right) + s(x).
\end{split}
\end{equation}
Assuming that the initial size is constant, the steady state solution is given by the solution of the inhomogeneous linear equation:
\begin{equation}
K - s_{C} \theta(x-x_{0}) = -f_{0} \rho + \sigma^2 \rho' + e^{-x} \frac{\mu + \nu}{2} \rho'.
\end{equation} 
The full solution is the sum $\rho=\rho_0+\rho_1$ of the particular solution,
\begin{equation}
\rho_0(x) = 
\begin{cases}
\frac{K}{|f_{0}|} {\rm \  for\  }x < x_{0} ,\\
\frac{K-s_{C}}{|f_{0}|}{\rm \  for \  }x >x_{0},
\end{cases}
\end{equation}
and the solution $\rho_1$ to
the homogeneous equation
\begin{equation}
 f_{0} \rho_1 = \sigma^2 \rho_1' + e^{-x} \frac{\mu+\nu}{2} \rho_1'
\end{equation}
of solution:
\begin{equation}
\rho_1(x) =   K'  \left[  \frac{ e^{x} + \frac{(\mu+\nu)}{2 \sigma^2}  }{ 1+  \frac{(\mu+\nu)}{2 \sigma^2}  } \right]^{- \alpha},
\end{equation}
with $\alpha=|f_{0}|/ \sigma^2$.
Therefore, for $x>x_{0}$
\begin{equation}
\rho(x) =    K'  \left[  \frac{ e^{x} + \frac{(\mu+\nu)}{2 \sigma^2}  }{ 1+  \frac{(\mu+\nu)}{2 \sigma^2}  } \right]^{- \alpha} +   \frac{K-s}{|f_{0}|}
\end{equation}
we set $K=s$ for convergence and obtain the steady state clone size distribution for large $x$
\begin{equation}
\rho(x) =    \left[  e^{x} + \frac{\mu+\nu}{2 \sigma^2}   \right]^{- \alpha} ,
\end{equation}
or in terms of the clone size
\begin{equation}
\rho(C) =    \frac{1}{C \left(  C + \frac{\mu+\nu}{2 \sigma^2}   \right)^{ \alpha} }.
\end{equation}
We see that  the white noise solution with birth--death noise has the same large clone power law behaviour as without birth--death noise. Fig.~\ref{bdnoeff} illustrates how birth death noise in the clone-specific fitness models with time correlated noise also does not affect the power law exponent but only the cut off of the power law.

\section{Data analysis}
\label{dataanalysis}

In the main text we report values of the power law exponents and power law cut off values obtained from the high throughput sequencing repertoire study of clone size distributions of zebrafish B-cell heavy chain receptors of Weinstein et al. \cite{quake-2009}.  We extracted the power law exponent and the best fit for the starting point of the power law, defined as its lower bound cutoff, from the discrete clone size distributions plotted in Fig.~1 of the main text using the methods discussed by Clauset and Newman \cite{newmanclauset}. Specifically, for each point of the cumulative clone size distribution we compute an estimate of the power law exponent with that point as cutoff (i.e the best
fit of the power law including only the values of the distribution above that point) using
\beq\label{alphaexp}
\alpha(C_{\rm min}) = 1 + n \left[ \sum_{i=1}^n  \log \left(\frac{C_{\rm i}}{C_{\rm min}}  \right)  \right],
\eeq
where $C_{\rm min}$ is the cut off and $n$ is the number of points with y-axis values above $C_{\rm min}$. For each of these cut-off values we compute the Kolmogorov-Smirnov distance between the data and the estimated power law distribution:
\beq
d(C_{\rm min}) = {\rm max}_{C>C_{\rm min}} \abs{ F_{d}(C) - F_{e} (C;C_{\rm min})  }
\eeq
where the maximum is taken over all values above the cut off $C_{\rm min}$, $F_{d}$ is the cumulative distribution function (CDF) of the data and $F_{e}(C;C_{\rm min})$ is the CDF of the estimated power law distribution with $C_{\rm min}$ as a cutoff, using Eq.~\ref{alphaexp}. The the cut off is taken to be the minimum of this distance over all possible cut off values and
the exponent is the exponent found for this value. 

The obtained power law parameters are presented in Table~\ref{Table1}. The power law exponent gives reproducible values for different individuals and agrees with values of the same exponent obtained from human data \cite{Bolkhovskaya2014}. We note that the power law exponent of the cumulative distribution function is $\alpha$ for a power law distribution with exponent $1+\alpha$. As discussed in detail in the main text, the reliability of the cutoff estimate $C^*$ is sensitive to experimental precision of capturing the rare clones. In the presented dataset the reads were not barcoded and the counts had to be renormalized by a known PCR amplification factor. Therefore, these normalized counts could not to used as normal counts, making the definition of a cut-off clone size problematic. To overcome this problem, we estimate the power law cut-off from the value of the cumulative distribution function at  the cut-off clone size (instead of the cut-off clone size itself). That value is invariant under rescaling of absolute clone size values, unlike $C^*$.

We notice that the steady state solution is invariant under a full rescaling of time in the equations of the dynamics. This means that the system can be described by two dimensionless parameters, $\alpha = f_0 \lambda^2/\gamma^2$ and $\lambda^3/\gamma^2$, and the introduction size $C_0$. Fitting $\alpha$ to data and assuming value for $C_0$, we can compare the value of the power law cut-off in data and in simulations to fit the remaining dimensionless parameter, $\lambda^3/\gamma^2$. Estimating $f_0$ based on thymic output we can predict the order of magnitude of $\lambda$ and $\gamma$.

\begin{table}
\begin{center}
\begin{tabular}{ c | c | c  | c}
  {Fish} & $1+ \alpha$ & $C^{*}$ &  $\log(1-\mathrm{CDF}(C^*))$ \\
\hline
  {A} & 2.0591 & 32.6445  & - 3.1389 \\
  {B} & 2.0214 & 10.7231 & -1.8644 \\
  {C} & 2.0708 & 16.7386 &  -2.4655 \\
  {D}&  2.0670  & 14.9313 &    -2.1492 \\
  {E} & 2.0529 & 8.2685 &    -1.8332   \\
  {F} & 2.0006 & 5.8972 &  -1.6161  \\
  {G} & 1.9867  & 52.2909 &   -2.7329 \\
  {H} &  2.2242 &    32.1719 &   -2.6877  \\
  {I} &  2.0835 & 18.4385 &   -2.2757  \\
  {J} & 1.6907 &    44.4885 &  -2.2877 \\
  {K} &     1.7641 & 3.6030 & -0.9907  \\
  {L} & 1.9417 & 18.5298 &      -2.2730  \\
  {M} & 1.9901 & 18.5531 &       -2.2031  \\
  {N} &     1.8877 & 108.4732 &  -2.7984 \\
\end{tabular}
\end{center}
  \caption{Fit of the power law exponent of the clone size distribution $1+\alpha$ and power law cut-off value $C^*$ for zebrafish B-cell heavy chain D segment data from Weinstein et al \cite{quake-2009} presented in  Fig.~1. The fit for 14 fish (named A to N) shows a similar fit of the power law exponent. }
   \label{Table1}
\end{table}

\section{Cell specific simulations}
\label{cellspecificnoise}
\begin{figure}
\includegraphics[width=0.5\textwidth]{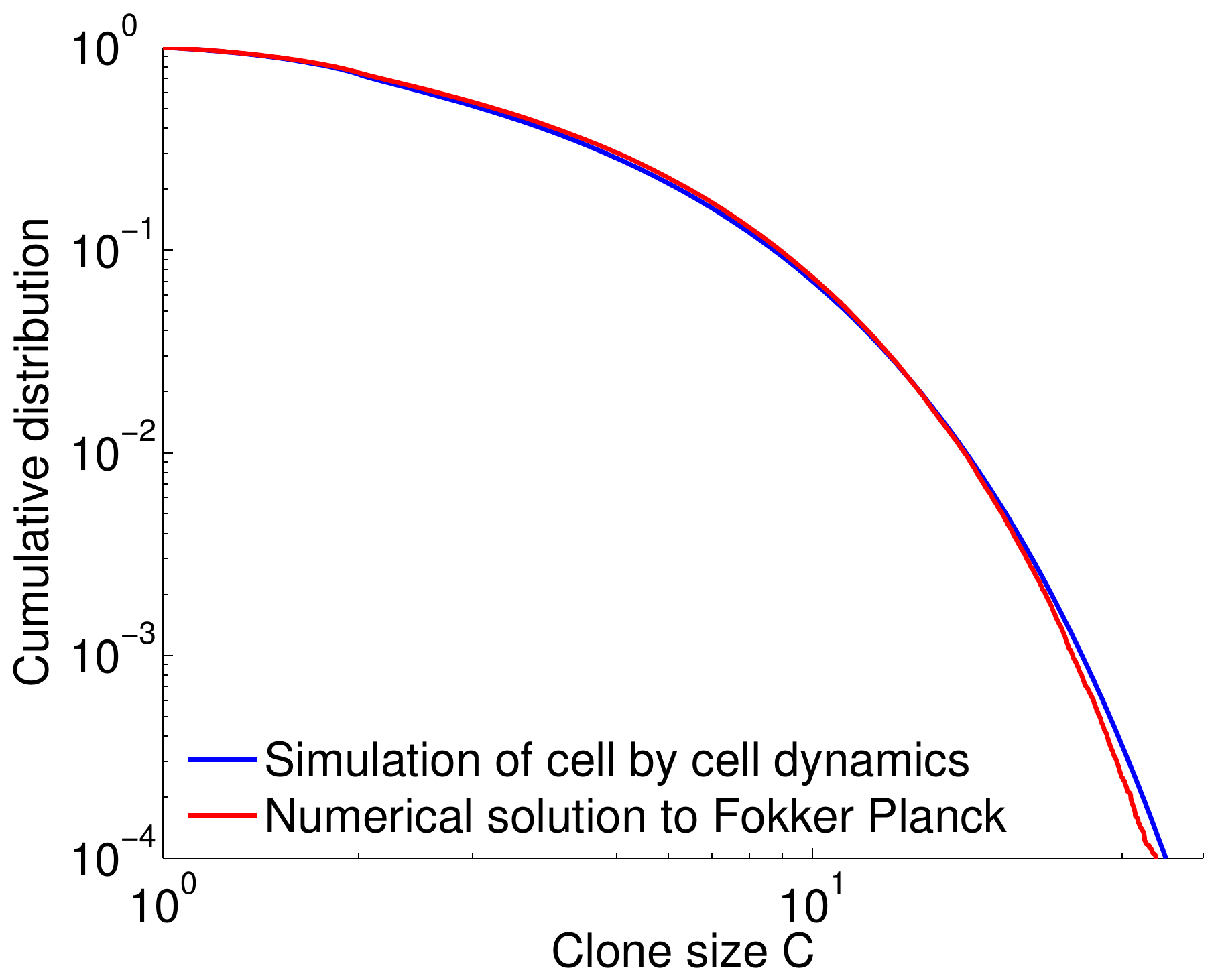}
\caption{Comparison of the Fokker-Planck solution (red line) and explicit Gillespie simulations of the dynamics (blue line) for the cell specific fitness model discussed in the ``A model of fluctuating phenotypic fitness'' Results section of the main text, show good agreement allowing us to use the population level Fokker-Planck solution to explore parameter space. Parameters were taken to be $\mu = 0.5$ day$^{-1}$, $\nu=0.8$ day$^{-1}$, $C_0 = 2$, $\lambda = 4$ days$^{-1}$ and $\gamma_c=4$ day$^{-3/2}$.  \label{il7example}}
\end{figure}

In the ``A model of fluctuating phenotypic fitness'' Results section of the main text, we present results of Fokker-Planck simulations for the cells dynamics. Here we verify that the stochastic dynamics of cells subject to a fluctuating cell-specific fitness are well approximated at the population level by a Fokker-Planck equation with a source term accounting for the import of new clones by comparing its numerical steady-state solution obtained by a finite elements method to explicit Gillespie simulations. We simulated the dynamics of clones using a Gillespie algorithm where cell division and death are accounted for explicitly and depend linearly on a fitness $f_c(t)$ fluctuating according to Eq.~7. The death rate is kept constant (above the average birth rate) and the fluctuations of the fitness only affect the birth rate (with the constraint that the birth rate is always positive). The agreement between the results of this detailed simulation and the Fokker-Planck solution, shown in Fig.~\ref{il7example}, validates the linear-noise approximation for the birth-death noise as well as the averaging argument leading to Eq.~8 and 9. This allows us to rely on the Fokker-Planck solution to explore parameter space.

\begin{figure}
\includegraphics[width=0.5\textwidth]{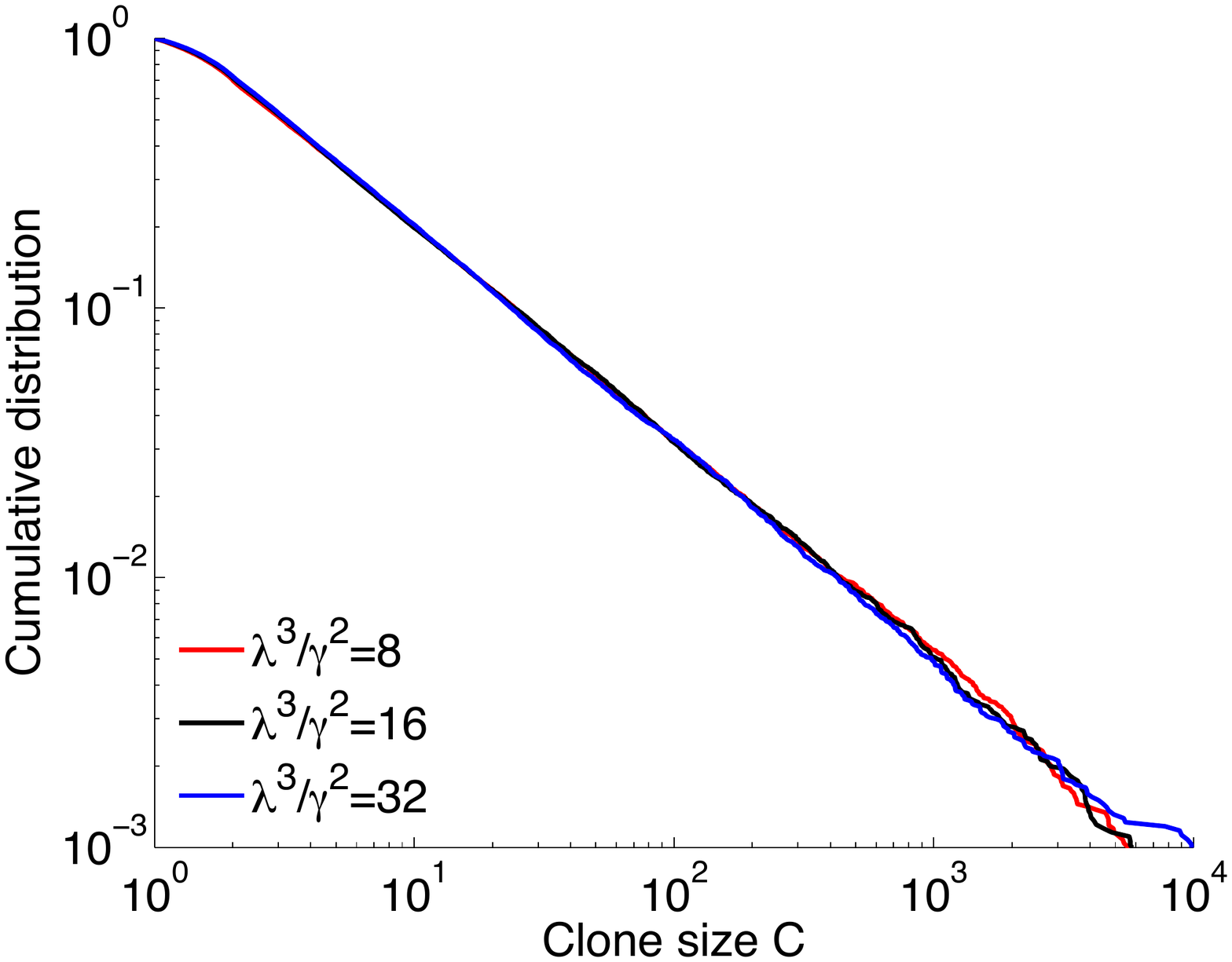}
\caption{Varying the dimensionless parameter related to the effective strength of antigen fluctuations relative to their characteristic lifetime $\lambda^3/\gamma^2$ does not affect the exponent of the power law if the ratio between exponential decay $\lambda$ and standard deviation of the variation $\gamma$ is kept constant. For all three curves the exponent is $\alpha = 0.8 $ and  $\mu = 0.5$ days$^{-1}$, $\nu=0.8$ days$^{-1}$, $C_0 = 2$ while  $\lambda$ and $\gamma$ vary. \label{exposame} }
\end{figure}

\begin{figure}
\includegraphics[width=0.5\textwidth]{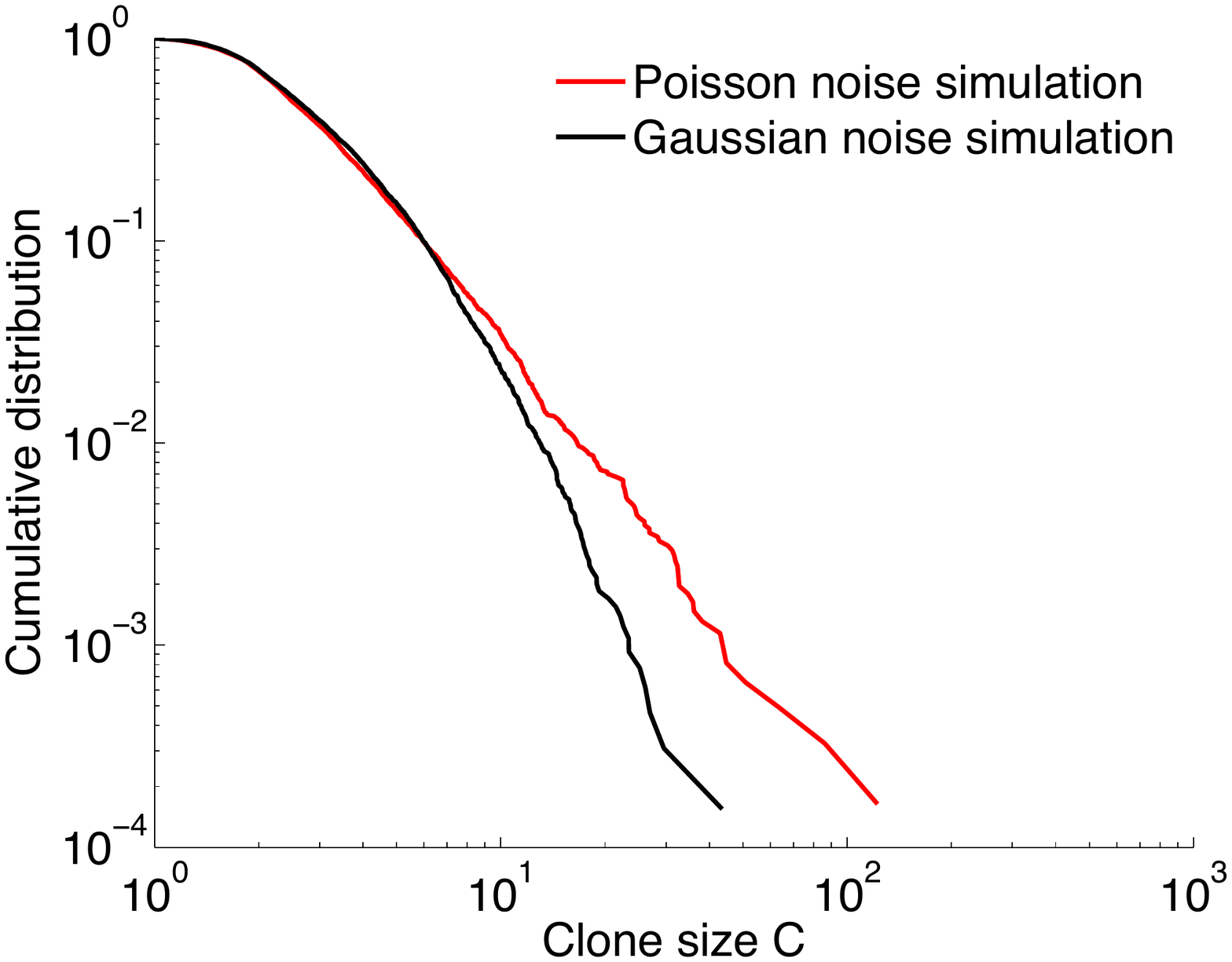}
\caption{Large deviations can influence the effect of Poisson noise on the simulated clone size distributions and create a discrepancy between Poisson noise (red line) and the Gaussian approximations (black line) we assume in the main text. The discrepancy is most apparent for small clones.
We simulated the Langevin dynamics of the Gaussian model with $\mu = 0.5$ day$^{-1}$, $\nu=1$ day$^{-1}$, $C_0 = 2$, $\lambda = 3$ day$^{-1}$ and $\gamma=1$ day$^{-3/2}$ and the same dynamics with Poisson noise and  $\mu = 0.5$ day$^{-1}$, $\nu=1$ day$^{-1}$, $C_0 = 2$, $\lambda = 3$ day$^{-1}$ and $s_{\rm A}=10^{7}$ day${}^{-1}$.  In both cases we introduce $s_C=2000$ new clones per day.  \label{poigau}}
\end{figure}

\section{Model of cell-specific fitness fluctuations, and its limit of no heritability}
\label{cellspecifi_nomemory}
The cell specific fitness model described in the ``A model of fluctuating phenotypic fitness'' Results section of the main text arises as a description of a population where each cell experiences its own growth fluctuations but cells deriving from the same lineage remain correlated. In this Appendix we derive the equations that describe the dynamics of clones in this system. 

Each cell $c$ experiences a time-correlated multiplicative noise from environmental growth factors. For cells $j$ in a given cell lineage (or clone) $i$, each individual cell's fitness follows the stochastic dynamics:
\begin{equation}
 \partial_{t} f_{c}(t) = - \lambda f_{c} + \sqrt2 \gamma_{c} \eta_{c}  \\
\end{equation}
where $\langle  \eta_{c}(t)  \eta_{c}(t') \rangle= \delta (t-t')$. Averaging over all cells in the clone, 
we obtain
\begin{equation}\label{eq:IL7d2}
\begin{cases}
\partial_{t} C_i = f_{0} C_i + f_i C_i +\sqrt{(\mu+\nu) C_i} \xi_i  \\
\partial_{t} f_i = - \lambda f_i + \sqrt{\dfrac{2}{C_i}} \gamma_c \eta_i,   \\
\end{cases}
\end{equation}
where $f_i$ is the average fitness in clone $i$
\begin{equation}
 f_i(t) = \frac{1}{C_i}\sum_{c\in i} f_{c}(t),
\end{equation}
and where we have added a birth-death noise term $\sqrt{(\mu+\nu)C_i} \xi_i$. We use the \^Ito convention for the birth-death noise, $\<\xi_i(t)\xi_i(t')\>=\delta(t-t')$ and the Stratanovich one for the environmental noise $\<\eta_i(t)\eta_i(t')\>=\delta(t-t')$. 
The equivalent equations for $x=\log C$ are
\begin{eqnarray}
    \partial_{t} x_i &= &f_{0}  + f_i  + \sqrt{\mu+\nu} e^{-x_i/2} \xi - e^{-x_i} \frac{\mu+\nu}{2} \\ 
    \partial_{t} f_i &= &- \lambda f_i +\sqrt{2}  e^{-x_i/2} \gamma_c  \eta_i
\end{eqnarray}
and the Fokker-Planck equation is
\begin{equation}
\label{fullfokil}
\begin{split}
\partial_{t} \rho(t,x,f)   =& - (f_{0} +f )\partial_{x} \rho + \lambda \partial_{f} (f \rho) + e^{-x} \gamma_c \partial_{f}^{2} \rho \\ &+ \frac{\mu+\nu}{2}  \partial_{x} (e^{-x} \rho  )   + \frac{\mu+\nu}{2} \partial_{x}^{2} ( e^{-x} \rho)\\
&+s(x,f),
\end{split}
\end{equation}
where $s(x,f)$ is the joint distribution of size and fitness or newly arriving clones (from thymic or bone marrow output).
This is the full Fokker-Planck equation that is solved numerically in the main text using the finite elements method.

Because of the $1/\sqrt{C_i}$ prefactor in front of the noise term, we could expect fitness fluctuations to behave like a birth-death noise in the limit of low heritability ($\lambda \rightarrow \infty$). 
In the remainder of this Appendix we show that this is not the case, and we show how to take the limit of no heritability properly.

Consider the limit of  $\lambda \rightarrow \infty$ and  $\gamma_c \rightarrow \infty$, keeping the ratio $\gamma_c / \lambda $ constant, so that $f$ does not become infinitesimally small. The equation for the environmental stimulation $f$ in $x = \log C$ space is given by (in Stratanovich convention)
\begin{equation}
\partial_{t} f =  - \lambda f + \sqrt{2} \gamma_{c} e^{-x/2} \eta .
 \end{equation}
Direct integration gives
\begin{equation}
f(t)   =   \sqrt{2} \gamma_c \int_{0}^{t} e^{-\lambda u } e^{-x( t-u )/2 } \eta( t-u ) du 
\end{equation}
and we divide the integral into two sub-integrals for $k>0$
\begin{equation}
\begin{split}
  f(t)  =  \sqrt{2} \gamma_c \int_{k/\lambda  }^{t} e^{-\lambda u } e^{-x( t-u )/2 } \eta( t-u ) du \\  + \sqrt{2} \gamma_c \int_{0}^{k/\lambda } e^{-\lambda u } e^{-x( t-u )/2 } \eta( t-u ) du. 
\end{split}
\end{equation}
With infinite precision, for any value of $t$, we set the integral of $\eta$ to be bounded and obtain the first integral is with probability $1- \epsilon $ smaller in norm than  
\begin{equation}
 \sqrt{2} \gamma_c \sqrt{t}  K(\epsilon) e^{- k  },   
\end{equation}
where $K(t)$ is a constant to control the variations of the integral of $\xi$ with probability $\epsilon$ (where $K(t, \epsilon)= \Phi ^{-1} ( 1- \frac{\epsilon}{2} ) $ with $\Phi$ the CDF of a standard Gaussian distribution).

The second sub-integral is
\begin{equation}
\begin{split}
 \sqrt{2} \gamma_c \int_{0}^{k/\lambda } e^{-\lambda u } e^{-x( t-u )/2 } \eta( t-u ) du \\ \approx e^{ -x(t^-)/2 } \eta(t)             \sqrt{2} \frac{\gamma_c}{\lambda} (1 - e^{ - k  }).   
\end{split}
\end{equation}
We choose $k = \sqrt{\lambda}$ and in the limit of $\lambda \rightarrow \infty$ and  $\gamma_c \rightarrow \infty$ keeping $\gamma_c / \lambda ={\rm const}$ we obtain the final form of environmental fluctuations 
\begin{equation}\label{limf}
 f(t) \longrightarrow \sqrt{2 \frac{\gamma_c}{\lambda}} e^{-x(t^-)} \eta(t),
\end{equation}
where $t^-$ means the left-hand limit. $f(t)$ depends only on the past, which means that in $x=\log C$ space the noise is similar to a birth-death noise in the \^Ito convention. Yet in terms of clone sizes $C$  additional \^Ito terms make the effect of environmental fluctuations  different from classical birth-death dynamics.

\section{Model solutions for cell-specific fitness fluctuations in the limit of no heritability}
\label{purecellspe}
In this Appendix we solve the model of cell-specific fitness fluctuations in the limit where trait  heritability is low. In this limit, the dynamics is described by a model with an instantaneous random fitness that is uncorrelated for cells in the same clone.
The resulting Langevin equation reads:
\begin{equation}
\label{neutrallim}
\frac{dC_i}{dt} = f_0 C_i + \sqrt{2C_i} \frac{\gamma_c}{\lambda} \eta_i +
\frac{\gamma_c^2}{\lambda^2}    + \sqrt{(\mu+\nu)C_i} \xi_i
\end{equation}
where all noise is treated in the \^Ito convention, and where the extra term $\gamma_c^2/\lambda^2$ comes the converting back the low-heritability limit of the fitness fluctuations, given by Eq.~\ref{limf}, into $C=e^x$ space. We note that although the fitness and birth-death noise have very similar forms, the birth-death noise is self-generated and intrinsic, while the fitness noise is environmental and extrinsic. This small difference greatly affects the steady-state clone size distribution.

\begin{figure}
\includegraphics[width=0.5\textwidth]{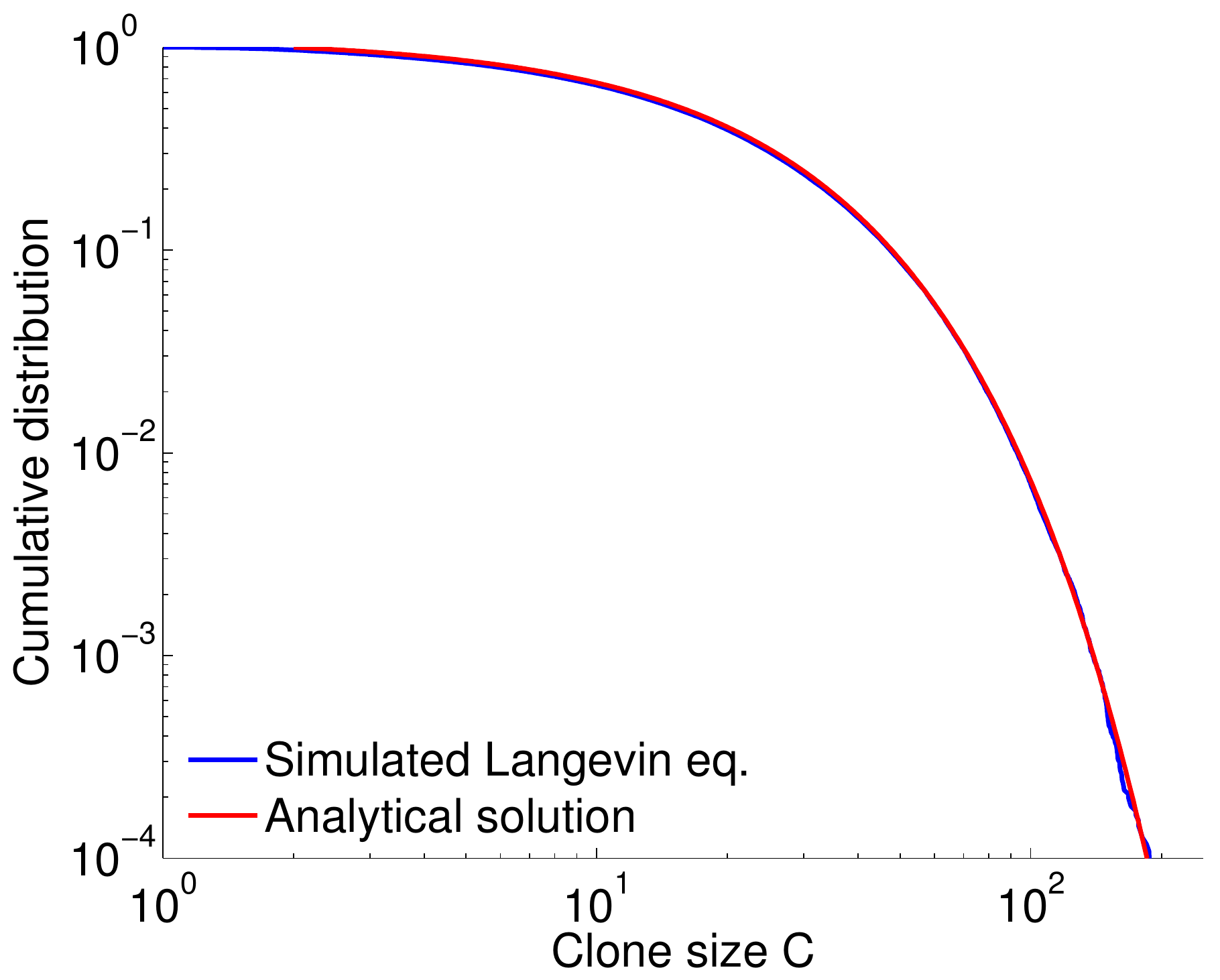}
\caption{The result of a simulation of the Langevin equation of the white noise cell-specific fitness model (blue line) compared to the analytical prediction of Eq.~\ref{whitenoisecell} (red line) show very good agreement. The parameters are $\mu = 0.2$ day$^{-1}$, $\nu=0.4$ day$^{-1}$, $C_0 = 2$, $\lambda = 4$ day$^{-1}$ and $\gamma_c=8$ day$^{-3/2}$. \label{ilonednobd}}
\end{figure}

\begin{figure}
\includegraphics[width=0.5\textwidth]{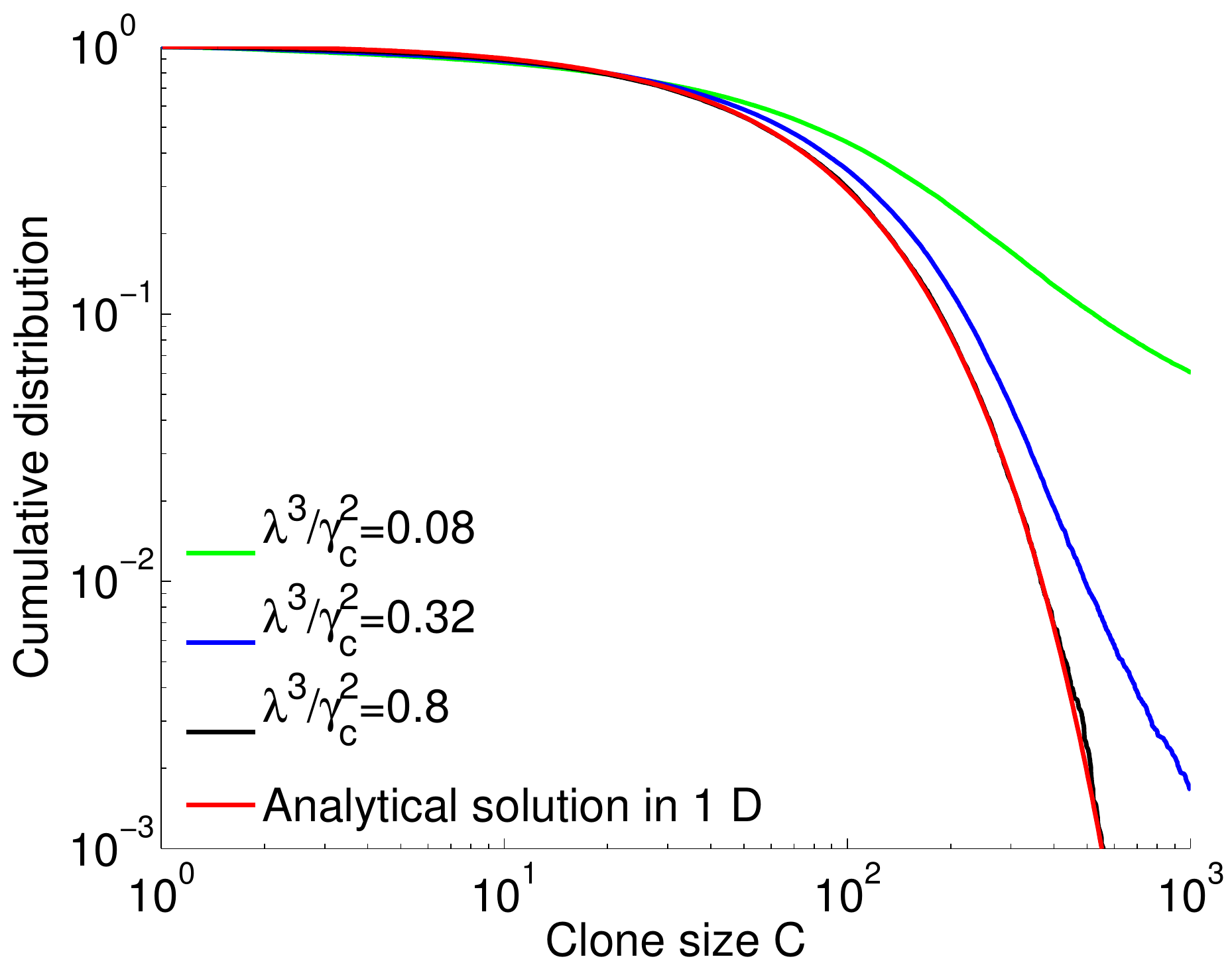}
\caption{Convergence of the cell-specific fitness models (Eq.~\ref{fullfokil}) without birth-death noise to Eq.~\ref{whitenoisecell} in the limit of no heritability ($\lambda\to \infty$).  For all four curves $\alpha = 0.2$. Parameters used: $\mu = 0.2$ day$^{-1}$, $\nu=0.25$ day$^{-1}$, $C_0 = 2$ and $1000$ new clones  introduced each day.
\label{ilconvnobd}}
\end{figure}

To see this, we first consider the case of  no birth-death noise. In the cell-specific fitness model consider the following equations with the Stratanovich rule: 
\begin{equation}
\begin{cases}
    \partial_{t} C_i = f_{0} C_i + f C_i  , \\ 
    \partial_{t} f_i = - \lambda f_i + \sqrt{\frac{2}{C_i}} \gamma_c  \eta_i ,
\end{cases}
\end{equation}
and its equivalent for $x=\log(C)$
\begin{equation}
\begin{cases}
    \partial_{t} x_i = f_{0}  + f_i ,   \\ 
    \partial_{t} f_i = - \lambda f_i + e^{-x_i/2} \gamma_c  \eta_i 
\end{cases}
\end{equation}
In Appendix~\ref{cellspecifi_nomemory} we have shown that in the limit of $\lambda \rightarrow \infty$ and $\gamma_c\rightarrow \infty$, the system reduces to the one dimensional equation 
\begin{equation}
\partial_{t} x_i = f_{0} + e^{-x_i/2} \sqrt{2} \frac{\gamma_c}{\lambda} \eta_i
\end{equation}
with the \^Ito rule for the white noise $\eta_i$. 
The corresponding Fokker-Planck equation is 
\begin{equation}
\partial_{t} \rho = \partial_{x} (- f_{0} \rho) + \frac{1}{2} \partial^{2}_{x} \left[ \frac{2 \gamma_c^2}{\lambda^2} e^{-x} \rho  \right] + s(x).
\end{equation}
Assuming a deterministic introduction size $s(x)=s_C \delta(x-x_{0})$, at steady-sate we get  
\begin{equation}
K - s_C \theta (x-x_{0})= -f_{0} \rho+ e^{-x} \frac{\gamma_c^2}{\lambda^2} \rho' -  \frac{\gamma_c^2}{\lambda^2} \rho e^{-x},
\end{equation}
which for  $x>x_{0}$ is solved by
\begin{eqnarray}
\rho (x) =   e^{-  e^{x} / C_m +x} \Big[ K Ei(e^{x} / C_m) - K Ei(C_m^{-1}) \\ -\frac{s_C \lambda^2}{\gamma_c^2} Ei(\frac{ e^{x}}{C_m})  +  \frac{s_C \lambda^2}{\gamma_c^2} Ei(\frac{e^{x_{0}}}{C_m}) \Big] ,
\end{eqnarray}
where $K$ is an integration constant, $Ei$ is the exponential integral function and
\begin{equation}
C_m= \frac{\gamma_c^2}{|f_0| \lambda^2}.
\end{equation}
The divergence of $Ei$ at infinity sets $K={s_C\lambda^2}/({\gamma_c^2})$ and the clone size distribution is
\begin{equation}
 \rho(x) =
 \begin{cases}
  \left(Ei(e^{ x}/C_m)- Ei(C_m^{-1})\right)  e^{  - e^{x} /C_m+ x } {\rm \    for \   } x<x_{0}  \\ 
   \left(Ei(  e^{x_{0}} / C_m)- Ei(C_m^{-1})\right)  e^{- e^{x} C_m + x }  {\rm \  for \   } x>x_{0}
 \end{cases}
\end{equation}
or in terms of $x=\log C$
\begin{equation}
\label{whitenoisecell}
 \rho(C) =
 \begin{cases}
e^{-C/C_m} \left(Ei(C/C_m)- Ei(C_m^{-1})\right) {\rm \    for \   }C<C_{0}  \\ 
 e^{-C/C_m} \left(Ei( e^{x_{0}}/C_m)- Ei(C_m^{-1})\right)  {\rm \  for \   }C>C_{0}
 \end{cases}
\end{equation}
The validity of this solution is checked in Fig.~\ref{ilonednobd} and the convergence of the full solution of Eq.~\ref{fullfokil} (with no birth-death noise) to the analytical solution in the limit of no heritability ($\lambda\to\infty$) is show in Fig.~\ref{ilconvnobd}.

For comparison, in a pure birth-death process (no fitness fluctuations) the clone-size distribution is, for $C$ large enough, $\rho(C) \sim {e^{- C  / C_m}}/{C} $ where $C_m = {(\mu+\nu)}/(2(\mu-\nu))$, as shown in Appendix \ref{appbirthdeath}. These two solutions both have an exponential cutoff, but have very different power-law exponents, corresponding to $\alpha=0$ and $\alpha=-1$, respectively.

\begin{figure}
\includegraphics[width=0.5\textwidth]{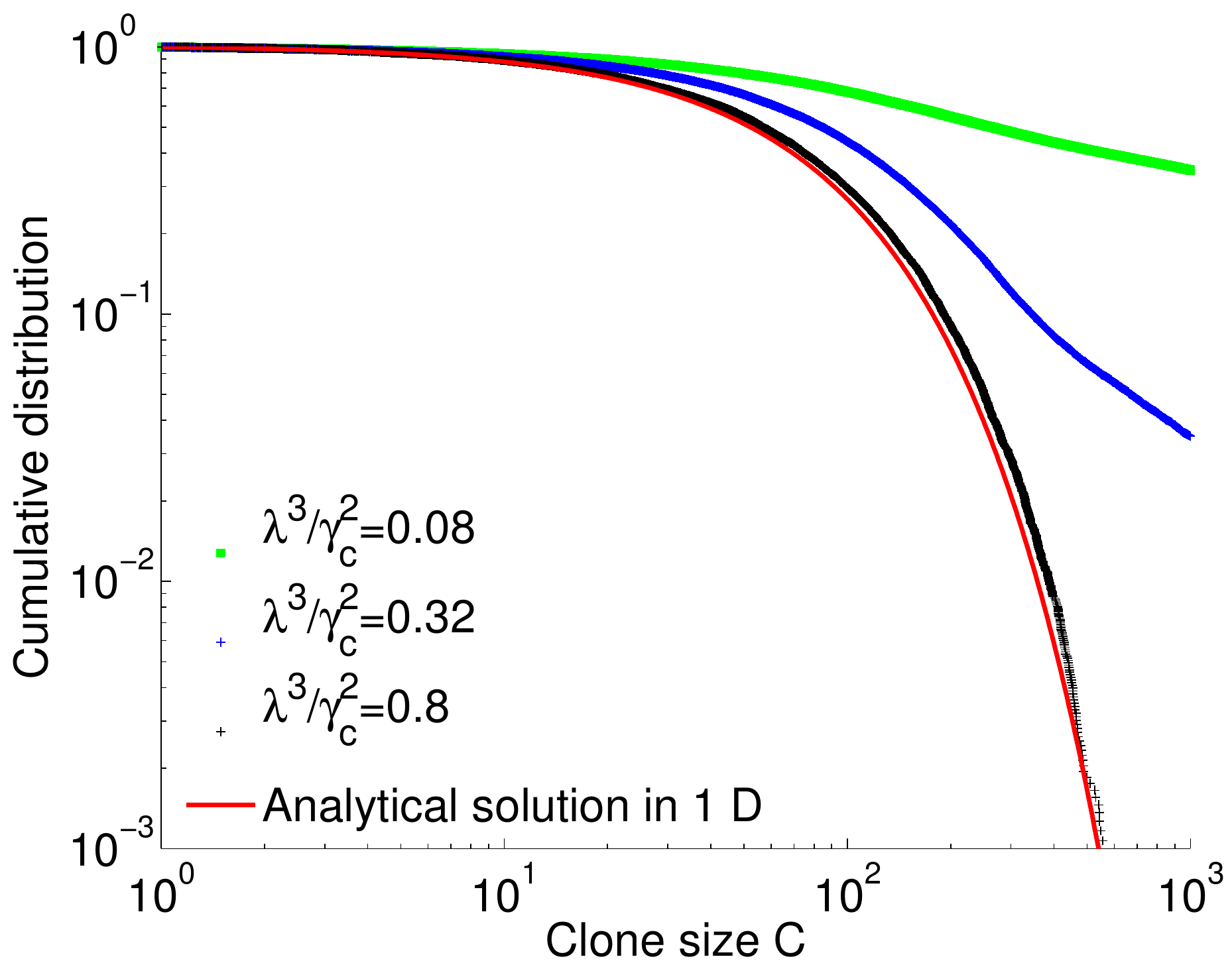}
\caption{Convergence of the cell-specific models (Eq.~\ref{fullfokil}) with birth-death noise to the analytical result of Eq.~\ref{soluneutlim} (red line). Keeping constant $\alpha$ while $\lambda \rightarrow 
\infty$ and $\gamma_c  \rightarrow  \infty$ we recover the solution of Eq.~\ref{soluneutlim}. Parameters are the same as in Fig. \ref{ilconvnobd}  \label{ilconvbd}}
\end{figure}

We now add the birth-death noise, {\em i.e.} consider both types of noise, still in the limit of no heritability. The corresponding Langevin equation reads:
\begin{equation}
\partial_{t} x_i = f_{0}   + \sqrt{\mu+\nu} e^{-x_i/2} \xi - e^{-x_i} \frac{\mu+\nu}{2} + e^{-x_i/2} \frac{\sqrt{2} \gamma_c}{\lambda} \eta
\end{equation} 
where all noise is in the  \^Ito convention. 
Integrating the Fokker Planck associated to this equation gives at steady state condition
\begin{equation}
K - s_C \theta(x-x_{0}) = - f_{0} \rho  + \left[ \frac{\mu+\nu}{2} + \frac{\gamma_c^2}{\lambda^2} \right] e^{-x} \rho ' - \frac{\gamma_c^2}{\lambda^2} e^{-x} \rho.
\end{equation}
In order for $\rho$ to be well defined we set $K = s_C$. For $x>x_{0}$ the equation is homogeneous and solved by separation of variables:
\begin{equation}
\frac{d \rho}{\rho} e^{-x} \left[ \frac{\mu+\nu}{2} + \frac{\gamma_c^2}{\lambda^2}   \right] = \left( f_{0}  + \frac{\gamma_c^2}{\lambda^2}  e^{-x}  \right) \rho,
\end{equation}
and gives the solution:
\begin{equation}
\label{soluneutlim}
\rho(C)  =  \frac{K e^{ -  C/C_m}}{C^{1 + \alpha}},
\end{equation}
with 
\begin{equation}
 \alpha = -{\left({1+ \frac{(\mu+\nu) \lambda^2}{2 \gamma_c^2}}\right)}^{-1},
\end{equation}
which is a power-law with an exponent $0\leq 1+\alpha\leq 1$ and an exponential cutoff
\begin{equation}
C_m= (\mu - \nu)^{-1} \left( \frac{\mu+\nu}{2} + \frac{\gamma_c^2}{\lambda^2}  \right).
\end{equation}
The convergence of the solution of the full system, Eq.~\ref{fullfokil}, to this solution is checked in Fig.~\ref{ilconvbd}.

\bibliographystyle{pnas}
\bibliography{references}

\end{document}